\documentclass[hyper,a4paper,oneside]{JHEP3}

\usepackage{amsmath,amscd}
\usepackage{amsfonts,amssymb}
\usepackage{graphicx,shortvrb}
\usepackage{epsfig}
\usepackage{subfigure}
\makeatletter
\renewcommand{\subsubsection}{\@startsection{subsubsection}{3}{0mm}{-\baselineskip}{0.5\baselineskip}{\normalfont\normalsize\it}}
\makeatother

\def\s{\sigma}

         \def\pd{\partial}

\newcommand{\idn}{{1\relax{\kern-.35em}1}}
\newcommand{\Cf}{\mathbb{C}} \newcommand{\Rf}{\mathbb{R}}
\newcommand{\Zf}{\mathbb{Z}} 

\newcommand{\ads}[1]{$AdS_{#1}$}
\newcommand{\tads}[1]{$TAdS_{#1}$}

\newcommand{\SU}[1]{\mathop{\rm SU}(#1)}
\newcommand{\U}[1]{\mathop{\rm U}(#1)}
\newcommand{\SL}[1]{\mathop{\rm SL}(#1)}

\newcommand{\csp}[1]{\hspace{#1em},\hspace{#1em}}
\newcommand{\gsp}[2]{~#2\hspace{#1em}}
\newcommand{\rar}[1]{\hspace{#1em}\Rightarrow\hspace{#1em}}

\newcommand{\abs}[1]{\left\vert#1\right\vert}
\newcommand{\pdf}[2]{\frac{\partial #1}{\partial #2}}
\DeclareMathOperator{\tr}{Tr}  \DeclareMathOperator{\re}{Re} \DeclareMathOperator{\im}{Im}

\preprint{\small WIS/06/07-JUN-DPP}
\title{Thermal $AdS_3$, $BTZ$ and competing winding modes condensation}
\author{Micha Berkooz, Zohar Komargodski, and Dori Reichmann  \footnote{micha.berkooz,~
zkomargo,~dor.reichmann~@weizmann.ac.il}\\
Department of Particle Physics,\\The Weizmann Institute of Science,\\ Rehovot 76100, Israel }
\date{\today}
\abstract{We study the thermal physics of \ads3 and the $BTZ$ black
hole when embedded in String theory. The exact calculation of the
Hagedorn temperature in \tads3 is reinterpreted as the appearance of
a winding tachyon both in \ads3 and BTZ. We construct a dual framework for analyzing the
phases of the system. In this dual framework, tachyon condensation
and geometric capping appear on the same footing, bridging the usual
gap of connecting tachyon condensation to modifications of geometry.
This allows us to construct in a natural way a candidate for the
unstable phase, analogous to a small black hole in higher
dimensions. Additional peculiar effects associated with the Hagedorn
temperature and the Hawking-Page transition, some to do with the
asymptotic structure of $AdS_3$ and some with strong curvature
effects, are analyzed and explained. }
\keywords{Black holes in String Theory, Conformal Field Models in String Theory, Tachyon Condensation }
\begin{document}
\setcounter{tocdepth}{2}

\section{Introduction and summary of results}\label{z-tit-int}

The thermodynamical interpretation of classical gravity in
asymptotically \ads{d} spaces
\cite{Maldacena:1997re,Witten:1998qj,Gubser:1998bc,Witten:1998zw}
predicts a maximal temperature for a gas of thermally excited
strings (the Hawking-Page temperature \cite{Hawking:1982dh}). At
temperatures higher than the HP temperature the gas of strings
collapses into a black hole via a first order phase transition. If
we keep the gas of strings in an overheated metastable phase
(preventing the HP phase transition) while raising the temperature,
we will eventually reach the Hagedorn temperature where the gas of
strings must collapse into a black hole because the barrier
disappears and a tachyonic mode appears
\cite{Barbon:2001di,Barbon:2002nw,Barbon:2004dd,Horowitz:2006mr}. This tachyon is similar to the
Atick-Witten tachyon \cite{Atick:1988si} in flat space, which is
associated with the Hagedorn temperature for that configuration.

In this paper we explore both the Atick-Witten and Hagedorn phase
transitions in asymptotically \ads3 spaces using the exact, in
$\alpha'$, worldsheet description (for the most part). For both
thermal \ads3 and Euclidean $BTZ$ the boundary of spacetime looks
like a radial direction$\times S^1_t\times S^1_\theta$ where $t$ is
the Euclidean time direction and $\theta$ an angular variable. In
\tads3 the radial direction and $S^1_\theta$ combine to form a
2-dimensional disk and in the Euclidean $BTZ$ the radial direction
pairs up with $S^1_t$. Above the Hagedorn temperature a winding mode
tachyon appears around the $t$ direction. It is generally believed,
and has been argued in various ways (for example
\cite{Barbon:2001di,Barbon:2002nw,Barbon:2004dd,Horowitz:2006mr,Adams:2005rb,Silverstein:2006tm,Horowitz:2005vp})
that such a tachyon causes the $t$ circle to pinch, changing the
topology of the background to that of $BTZ$, to which the system
then relaxes.

This is a compelling scenario, but following this topology change in
details is rather complicated, as one needs to follow the flow of
the worldsheet through this topology change in which the tachyon
mixes with, or induces, metric deformations. What is sometimes done
for this case is either to argue the effects of the tachyon based on
general worldsheet RG properties, or to describe the two topologies
using a Ricci flow, which either ends or start from a singularity -
when the topology change occurs - and to glue the two flows at the
singularity in a somewhat ad-hoc (although correct) manner
\cite{Headrick:2006ti}.

There are many works on closed string tachyon condensation. It is
common to distinguish localized closed string tachyons and bulk
closed string tachyons. The former have a relatively mild effect as
they deform the geometry in the vicinity of the region in which they
are localized, usually capping the geometry and making a small part
of space disappear. This process may be followed by emission of some
perturbative massless and massive bulk modes. For references on the
subject of closed string tachyon condensation see
\cite{Horowitz:2006mr,Adams:2001sv,Harvey:2001wm,localized tachyons}
and references therein.\footnote{For a review of tachyon
condensation in a cosmological context see \cite{Berkooz:2007nm} and
\cite{applicationtimedep}. Phase transitions to bubbles of nothing
in \ads5 which are driven by winding tachyons  were discussed in
\cite{Balasubramanian:2005bg,He:2007ji}.} Bulk tachyons, on the other hand, are much harder
to understand, and their condensation is expected to reduce the
number of space time dimensions \cite{bulk tachyons}, as a
consequence of the Zamolodchikov c-theorem \cite{Zamolodchikov:1986gt}. The tachyons
that are relevant for phase transitions in $AdS$ spaces are either
localized tachyons or de-localized for $AdS_3$ with NS-NS fields
(which nevertheless have effects similar to localized tachyons).

In this paper we take some steps towards improving the understanding
of the flow around the topology change point. The immediate context
which we will discuss is the Hawking-Page and Hagedorn phase
transitions in \ads{3}, where we can rely on the notion of a dual
CFT, but we expect that a similar set of tools will be useful to
discuss topology change in more general cases. The main idea is to
convert the background geometric data into a tachyon condensation
problem, like in the FZZ duality \cite{FZZ}. The problem then
reduces to a simpler problem of comparing the strength of the
"geometric" tachyon wall with the strength of the new
Atick-Witten-like tachyonic wall. This can be made very precise in
the case of \tads{3}/$BTZ$ which is what we will do next.

Strings propagating in \ads3 are described by the $\SL{2,\Rf}$ WZW
model. To describe the thermal gas of strings we use the Euclidean
version of the CFT (the $H_3^+$ model) and compactify the Euclidean
time coordinate. By a sequence of T-dualities we bring this space to
the form of a $cigar\times S^1$ which is simply the coset
\begin{equation*}
    \frac{\SL{2,\Rf}}{\U1}\times \U1
\end{equation*}
(with some important gluing conditions). By applying an FZZ duality
the theory is mapped into a $sine-Liouville\times S^1$, which is a
$linear\ dilaton\times S^1\times S^1$ with an interaction term
coupling the Liouville mode ($\phi$) with a winding mode around the
circle related to a spatial angle $x$ which is roughly the angular direction of \ads3,
\begin{equation*}
    e^{b\phi}\cos R_x(x_L-x_R).
\end{equation*}

On the other hand, the Atick-Witten thermal tachyon is a winding
mode around another circle $\varphi$, which is roughly the Euclidean
time direction $t$, and it is of the form
\begin{equation*}
    e^{b'\phi}\cos R_\varphi(\varphi_L-\varphi_R)
\end{equation*}
This achieves the goal of putting both metric and tachyon
deformations on the same footing in a manifest way, which simplifies
considerably the analysis of the topology change point. For example,
the phase transition between the thermal gas and black hole is a
competition (by RG flow) between two sine-Liouville interaction
terms, both including winding modes but on different circles. More
generally, this method provides us with a ``unified" description of
the various backgrounds in which phases and phase transitions of
different kinds are treated similarly. These include the Hawking
Page phase transition, the Hagedorn phase transition and the fixed
points that exist between them.

At the end of the day, this dynamics is controlled by computing
dimensions of some specific operators in sine-Liouville theory
(which we review in \S\ref{y-tit-comp-fzz}). In the case that these
operators are irrelevant, they drive a geometric capping (and a new
CFT) by turning them on with a large coefficient which corresponds
to a first order phase transition in spacetime. The other case, when
they are marginal or relevant, is the case beyond the Hagedorn
temperature, where their condensation is exactly the Atick-Witten
tachyon condensation.

Another result of our investigations is the surprising understanding
that thermal \ads{3} and the $BTZ$ black hole at the same
temperature have a pathological canonical thermodynamic descriptions
if embedded in String theory (unlike the case in pure gravity).
However, they are expected to have well defined micro-canonical
description.

The paper is organized as follows:  In \S\ref{z-tit-rev} we briefly
review known results concerning String theory on thermal \ads3.
\S\ref{z-tit-hag} is devoted to the calculation of the Hagedorn
temperature for \ads3, clarification of the phase diagram and
discussion of some special phenomena which take place at strong
curvature. \S\ref{z-tit-comp} and \S\ref{z-tit-mid} contain the main
results. In \S\ref{z-tit-comp} we explain the mapping of the thermal
theory into a $cigar\times S^1$. Then, applying FZZ duality, we map
the theory into $sine-Liouville\times S^1$ where it is manifest that
String theory treats equally geometric and tachyonic capping.
\S\ref{z-tit-mid} contains a discussion of the properties of the
unstable fixed point (CFT) which separates the black hole and
thermal \ads{} phases, as well as a discussion of the possible flows
in the system. We conclude in \S\ref{z-tit-open} with a discussion
of open questions and propose directions for future research.

We would like to thank Shiraz Minwalla and Vadim Shpitalnik for
collaboration at early stages of this work. After this work was
completed, we received a draft of \cite{Lin:2007gi} which overlaps
with some of the results in \S\ref{z-tit-hag}, and \cite{Rangamani:2007ju}
which discusses the Lorentizan BTZ.

\section{Review}\label{z-tit-rev}

\subsection{Solutions of Euclidean gravity}\label{y-tit-rev-eucl}

The Euclidean \ads3 manifold is the 3 dimensional hyperbolic space
$\mathrm{H}_3^+$ (which is the analytic continuation of
$\SL{2,\Rf}$, and can be represented as $\SL{2,\Cf}/\SU{2}$), i.e
the space of Hermitian unimodular matrices
\begin{align}
    X = \begin{pmatrix}X_{-1}+X_1 ~&~ X_2+iX_3 \\ X_2-iX_3 ~&~ X_{-1} - X_1\end{pmatrix}
    &&  X_i\in\Rf
    && \det X =1.
\end{align}
The manifold has an $\SL{2,\Cf}$ isometry group acting on it as
$X\rightarrow AXA^+$ where $A\in\SL{2,\Cf}$. Throughout the paper
we use global coordinates for \ads3
\begin{equation}
    X = e^{\frac{i}{2}(-it+\theta)\s^2}e^{\rho\s^3}e^{\frac{i}2(-it-\theta)\s^2},
\end{equation}
where the $\s^i$ are Pauli matrices. The line element is given by
the square of the Maurer-Cartan form
\begin{equation}\label{TAdS3-geom}
    ds^2 =\frac{k}{2}\tr\left(X^{-1}dX\right)^2 = k\left(\cosh^2\rho\, dt^2 + d\rho^2 +\sinh^2\rho\,
    d\theta^2\right),
\end{equation}
where $k$ is the \ads3 radius squared.\footnote{throughout the
paper we use units in which $\alpha'=1$.} In the Euclidean theory,
one can make the coordinates $t$ and $\theta$ periodic,
\begin{align}\label{TADS-period}
    \theta \cong \theta+2\pi&&
    t+i\theta \cong t+i\theta-2\pi i\,\tau
    \quad;&&
    \tau \equiv i\frac{\beta}{2\pi}\left(1+i\mu\right),&
\end{align}
such that at the conformal boundary $\rho\rightarrow\infty$, the
geometry is a $\mathbb{T}^2$ with modular parameter $\tau$. We use
the name thermal \ads3 (or \tads3) for such a geometry with finite
$\tau$. The parameters $\beta$ and $\mu$ in \eqref{TADS-period} are
correspondingly the inverse temperature and chemical potential of
the canonical thermodynamic ensemble.

The $\mathrm{H}_3^+$ WZW CFT describing strings propagating in
\tads3 includes a constant dilaton (with arbitrary value) and an
imaginary B-field,
\begin{equation}
    H_{(3)} = dB_{(2)} = -2ik\sinh(2\rho) d\rho\wedge dt\wedge d\theta.
\end{equation}
The B-field is imaginary due to the analytic continuation from the
Lorentzian geometry where the B-field is real. The Euclidean
worldsheet action, on the other hand, is real due to the additional
factor of $i$ from the analytic rotation of the worldsheet time.

In gravity, the canonical ensemble contains all Euclidean gravity
solutions which have the same boundary geometry (including H-field)
up to diffeomorphisms. To find the classical GR solutions
contributing to the ensemble it is convenient to start from the
Lorentzian \ads3 (the $\SL{2,\Rf}$ group manifold) and construct new
gravity solutions by orbifolding with elements of the isometry
group. The interesting orbifolds are classified by conjugacy classes
of $\SL{2,\Rf}$, of which there are three types \footnote{The number
of conjugacy classes is infinite and labeled by the trace of the
$\SL{2,\Rf}$ matrix. It is convenient to divide them to hyperbolic,
parabolic, and elliptic $Tr(M)>=<2$ respectively.} which generate
three distinct types of orbifolds. The $\Zf$ orbifolds generated by
elements in the hyperbolic class are the $BTZ$ black holes
geometries. These are the only orbifolds which have a smooth
Euclidean continuation with finite parameter $\tau$. The various
orbifolds (by elements of $\SL{2,\Rf}$) form a complete
classification of solutions to 2+1 dim GR with constant negative
curvature \cite{Banados:1992wn,Banados:1992gq}.

The Euclidean $BTZ$ ($EBTZ$) solutions (which are $\Zf$ orbifolds of $\mathrm{H}_3^+$) are commonly expressed in
Schwarzschild coordinates \cite{Banados:1992wn,Banados:1992gq} with periodicities \eqref{TADS-period},
\begin{align*}
    &ds^2 = N^2 dt^2+N^{-2}dr^2+r^2\left(d\theta+N_\theta dt\right)^2\cr
    &N^2 = \frac{(r^2-r_+^2)(r^2+r_-^2)}{k^2\,r^2}\qquad
    N_\theta = \frac{r_+r_-}{r^2}&\cr
    &
    r_+ = -k\,\im\frac{1}{\tau}\qquad
    r_- = k\,\re\frac{1}{\tau}.
\end{align*}
The modular parameter $\tau$ is related to the Lorentizan $BTZ$ black hole mass and angular momentum via,
\begin{align}
    M = \frac{r_+^2+r_-^2}{k^2}=\abs{\frac1{\tau}}^2
    &&
    J=\frac{2r_+r_-}{k}=-\frac{k}{2}\im\left(\frac1{\tau^2}-\frac{1}{\bar\tau^2}\right).
\end{align}

Maldacena and Strominger \cite{Maldacena:1998bw} argued that,
starting from a \tads3 with modular parameter $\tau$, there exists
an $\SL{2,\Zf}$ family of solutions \footnote{Which can perhaps also
be constructed by using the technique described in this paper of
applying different sine-Liouville caps on different cycles of the
$\mathbb{T}^2$.} which at infinity have the modular parameter $\tau$.
The construction is as follows - start with the $\mathbb{H}_3^+$
manifold (Euclidean \ads3) orbifolded by a $\SL{2,\Cf}$ elements of
the isometry group, generating a \tads3 solution with modular
parameters $\tau$. The $\SL{2,\Cf}$ element generating this orbifold
is
\begin{align*}
    H = \begin{pmatrix}e^{i\pi\tau} & 0 \\ 0 & e^{-i\pi\tau}\end{pmatrix}.
\end{align*}
Consider another element of $\SL{2,\Cf}$ generating \tads3 with
conformal parameter $\tau'$ such that there is an $\SL{2,\Zf}$
transformation connecting these two modular parameters,
\begin{align}
    \tau' = \frac{a\tau + b}{c\tau +d}
    \quad;&&
    \begin{pmatrix}a & b \\ c & d \end{pmatrix}\in\SL{2,\Zf}.
\end{align}
Then, there exists a coordinate transformation which acts on all
coordinates, but in particular it performs an $\SL{2,\Zf}$
transformation near the boundary, changing the modular parameter
from $\tau'$ to $\tau$. Thus, for a given modular parameter $\tau$
which defines the ensemble, there is a family of Euclidean solutions
corresponding to elements of $\SL{2,\Zf}$. These solutions are part
of the same thermal ensemble. It is known that \tads3 with parameter
$\tau$ is diffeomorphic to the $EBTZ$ black hole with periodicity
$\tau'=-\frac1{\tau}$. This correspondence is a special case of the
construction outlined above. The nice feature about the latter
transformation is that it has a well defined continuation to
Lorentzian signature. The Euclidean action of the instantons was
calculated in \cite{Maldacena:1998bw},
\begin{equation}
    S_{instanton} = \frac{i\pi k}{2}\left(\tau'-\bar\tau'\right)
     = -\frac{2\pi^2k\beta}{c^2\beta^2+\left(c\beta\mu-2\pi d\right)^2}.
\end{equation}
Note that the action depends only on two of the three independent
paraments of $\SL{2,\Zf}$, indeed the transformation $\tau'=\tau+b$
does not change the geometry or the entropy.

At each value of $\tau$ there is a dominant phase given by the
lowest (negative) action instanton. At low temperature,
$\beta\rightarrow\infty$, the dominant saddle point is the thermal
\ads3 solution (i.e a gas of cold particles). At very high
temperature, $\beta\rightarrow 0$ the dominant saddle point has
$\tau' = -\frac{1}{\tau}$, which is diffeomorphic to the $EBTZ$
black hole with this (high) temperature $1/\beta$ and chemical
potential $\mu$. In the generic case there are various phases that
can dominate the path integral at intermediate value of the
temperature \cite{Dijkgraaf:2000fq,deBoer:2006vg}.

The simplest situation occurs for vanishing chemical potential
$\mu=0$, which is what we assume from now on in this paper. The
simplifications is that there are only two dominating phases, which
are the thermal \ads3 at low temperature and the $EBTZ$ black hole
at high-temperature. The phase transition occurs at the Hawking-Page
temperature where the Euclidean actions of both saddle points are
equal
\begin{equation}\label{HP-temp}
    \beta_{HP} = 2\pi.
\end{equation}

\subsection{The partition function}\label{y-tit-rev-part}

A comprehensive quantization of String theory on \ads3 background
was accomplished in
\cite{Maldacena:2000hw,Maldacena:2000kv,Maldacena:2001km}. In these
series of beautiful papers the authors analyzed the $\SL{2,\Rf}$ WZW
model (at level $k$) describing strings propagating in
$\mathrm{AdS}_3\times\mathcal{M}$, where $\mathcal{M}$ is an
internal CFT which makes the theory critical. The Hilbert space of
the $\SL{2,\Rf}$ WZW model is decomposed into representation of the
left/right-moving current algebra
$\widehat{\SL{2,\Rf}}\times\widehat{\SL{2,\Rf}}$. The unitary
representations contributing to the spectrum are pairs of continuous
representations
$\mathcal{C}_{j=1/2+is}^\alpha\times\mathcal{C}_{j=1/2+is}^\alpha$
and pairs of lowest/highest weight discrete representations
$\mathcal{D}_{j>1/2}^\pm\times\mathcal{D}_{j>1/2}^\pm$. This is an
oscillator expansion around geodesics with mass $m^2=j(j-1)$.  It is
known that these 3 representations form a complete basis for
$\mathcal{L}^2(\mathrm{AdS}_3)$ and are part of the Hilbert space of
the WZW model. In addition to the conventional representation
discussed above, the Hilbert space contains spectral flowed
representations
$\mathcal{D}_{j>1/2}^{\omega\pm}\times\mathcal{D}_{j>1/2}^{\omega\pm}$
and
$\mathcal{C^\omega}_{j=1/2+is}^\alpha\times\mathcal{C^\omega}_{j=1/2+is}^\alpha$
with $\omega\in\Zf$. These should be thought of as oscillator
expansion around long string. They furnish representations of the
spectral flowed algebra,
\begin{align*}
    \tilde J^\pm_n=J^{\pm}_{n\pm w}
    &&
    \tilde J^3_n=J^{3}_{n}-\frac{k}{2}w\delta_{n,0}.
\end{align*}
The reader is encouraged to read \cite{Maldacena:2000hw} for a complete description of the spectrum.

The $\SL{2,\Rf}$ theory at level $k$ has an interesting phase structure as a function of $k$, all the
representation discussed above exist only if $\frac12<j<\frac{k-1}{2}$. Thus, if $k<3$ there are no states in
the Hilbert space since the vacuum \footnote{The vacuum is assumed to be $\SL{2,\Rf}$ invariant. Hence, it has
vanishing Casimir $j(j-1)=0$ which implies $j=1$.} is projected out. A detailed study of the phase structure and
its physical interpretation was conducted in \cite{Giveon:2005mi} and references within.

The Euclidean \ads3 background is described by the $\mathrm{H}_3^+$~ coset WZW model (at level $k$). A canonical
description of the model uses the Poincar\'e patch coordinates for thermal \ads3,
\begin{equation}
    ds^2 = \frac{k}{y^2}\left(dy^2+dw d\bar w\right),
\end{equation}
where,
\begin{align}
    y = \frac{e^t}{\cosh\rho}
    &&
    w = \tanh\rho\,e^{t+i\theta}
    &&
    \bar w = \tanh\rho\,e^{t-i\theta}.
\end{align}
In \cite{Maldacena:2000kv} the partition function of the model was computed (following \cite{Gawedzki:1991yu})
\begin{multline}\label{partition}
    Z(\beta, \mu)=\frac{\beta(k-2)^{\frac{1}{2}}}{2\pi}\int_R\frac{d\tau
    d\bar{\tau}}{\sqrt{\tau_2}}e^{4\pi\tau_2\bigl(1-\frac{1}{4(k-2)}\bigr)}\sum_{h,\bar{h}}D(h,\bar{h})
    e^{2\pi i \tau(h+\bar h)}\cr
    \times\sum_{n,m}\frac{e^{-k\beta^2|m-n\tau|^2/4\pi\tau_2+2\pi(ImU_{n,m})^2/\tau_2}}{\abs{\vartheta_1(\tau,U_{n,m})}^2},
\end{multline}
where $D(h,\bar h)$ is the degeneracy of the internal \footnote{The internal CFT has central charge such that
$c_{int}+c_{\SL{2,\Rf}}=26$.} CFT, $\tau$ is the modular parameter of the worldsheet torus (not to be confused
with the spacetime torus used above), $R$ is a fundamental domain of the modular group $\SL{2,\Zf}$ and
\begin{equation*}
U_{n,m}(\tau)=\frac{i}{2\pi}(\beta-i\mu\beta)(n\bar{\tau}-m).
\end{equation*}
The partition function has poles in the modular parameter plane at $\tau_{pole} =\frac{r}{w}+i\frac{\beta}{2\pi
w}$ for integers $r$ and $w$. These appear due to the long strings in the spectrum \cite{Maldacena:2000kv}.

\subsection{Singular conformal field theory}\label{y-tit-rev-sing}

${AdS}_3\times S^3$ with NS-NS background is a solvable worldsheet
conformal field theory. It has, therefore, been a useful laboratory
to explore the AdS/CFT duality
\cite{Maldacena:1997re,Witten:1998qj,Gubser:1998bc,Aharony:1999ti}.
From the spacetime point of view, however, it is a rather
complicated system which does not conform to the standard ideas of
an ordinary field theory on the boundary. The associated pathologies
will manifest themselves in the thermal \ads{}/$BTZ$ discussion as
well.

The pathology that will concern us the most (other pathologies are related to it), is the fact that the
conformal field theory is "singular" \cite{Seiberg:1999xz}. The singularity is similar to the $R^4/\Zf_2$ target
space with $\theta=\int B_{S^2}=0$ where the integral is on the 2-cycle at the origin, and manifests itself in
the spacetime CFT as a continuum of operators with continuous scaling dimension. From the spacetime point of
view, these operators correspond to "long strings" - strings that stretch around $\theta$ direction in equation
\eqref{TAdS3-geom} at large values of $\rho$, which can continue to $\rho\rightarrow\infty$ if their energy is
above a finite threshold. Although the proper size of the $\theta$ angle increases, and the string stretches
more and more geometrically, this effect is canceled by the B field which permeates the string and the string
can still make it to infinity.

The rest of the paper will be devoted to the tachyons that take the
system between thermal \ads3 (\tads3) and the $EBTZ$ black hole.
Naively, and in all higher dimensions, \tads{} and Euclidean black
holes differ by normalizable modes at the boundary of \ads{}, and
therefore they are manifestly in the same ensemble. Correspondingly,
if there is a tachyon which takes us from one configuration to the
other, then it is localized in the interior of \ads{} and therefore
is a discrete state. This is not the case for \tads3 and $EBTZ$ in
the case that the background contains only NS-NS fields - the
tachyons that we will encounter (for zero angular momentum) are part
of a continuum, and match on to the long string spectrum at large
values of $\rho$.\footnote{After some kinematical "processing" that
will have to do with the difference between the Lorentzian space,
where the "long strings" are defined, and the Euclidean space, where
the interpolating tachyons are defined.} Since the interpolating
tachyons can make it all the way to the boundary, and their geometry
is different in \tads3 and $EBTZ$ (although related by $\SL{2,\Zf}$
of course), there is information on the boundary that distinguishes
the two backgrounds. Correspondingly, \tads3 and $EBTZ$ differ by
non-normalizable modes and including them in the same ensemble is
done by fiat.

Peculiar as it may sound, such pathologies are not unexpected in a
theory with long strings type "singularities", which are also
counter intuitive from the standard UV/IR perspective.\footnote{Not
to mention that these theories also exhibit (related) pathologies
even below the long string gap such as dimension zero operator which
is distinct from the identity (even on the short string sector)
\cite{Kutasov:1999xu,Giveon:2001up}.}

In \cite{Seiberg:1999xz} it was shown that one can go away from the
locus of singular spacetime CFTs by turning on some of the moduli of
the theory. In our case one goes away from the NS-NS ${AdS}_3\times
S^3$ background by turning on RR fluxes in the background. If the
scenario that we outlined in the paragraphs above is correct, then
this deformation should also change the continuum of interpolating
tachyons into a discrete set. Deformations of this type have been
constructed in \cite{Maldacena:1999mh,Dhar:1999ax} and it is easy to
see that this is indeed what happens.

We will refer to these background as "regulated", and discuss their
application to the \tads3/$EBTZ$ transition in
\S\ref{y-tit-comp-reg}. At this point we will just outline our
approach. We will work with the NS-NS background since this is the
only case that we can solve exactly, and will view the addition of
R-R fields as a regulator for the behavior near the boundary of
\ads3.  We are entitled to do so because in the regulated version
there is no additional pathology that arises from the vicinity of
the boundary.\footnote{The Philosophy is similar to the one
advocated in \cite{Adams:2001sv} where tachyons which are localized
on time like singularities are discussed. The tachyons resolve and
smooth the singularities but then expand outwards until, at infinite
time, they will reach the boundary of space. The fact that the
boundary conditions change between the initial worldsheet CFT and
the final worldsheet CFT (at infinite distance) is irrelevant for
the discussion of how the tachyons smooth the singularity.}

\section{Hagedorn temperature}\label{z-tit-hag}

On general grounds we expect String theory in thermal \ads3 to exhibit an exponential growth of the density of
states which leads to the breakdown of perturbative String theory at the Hagedorn temperature. In order to
compute the Hagedorn temperature we use Polchinski's trick \cite{Polchinski:1985zf} to replace the sum over
string windings $n$ in \eqref{partition} by the sum over copies of the fundamental domain. This is interpreted
as the summation over the spacetime field theory one loop amplitudes followed by a summation over the number of
particles. The resulting one-loop partition function is
\begin{multline}\label{partition-strip}
    Z(\beta,\mu)=\frac{\beta(k-2)^{\frac{1}{2}}}{8\pi}\int_0^{\infty}\frac{d\tau_2}{\tau_2^{3/2}}\int_{-1/2}^{1/2}d\tau_1
    e^{4\pi\tau_2\bigl(1-\frac{1}{4(k-2)}\bigr)}\sum_{h,\bar{h}}D(h,\bar{h})e^{2\pi i\tau (h+\bar h)}\cr
    \times\sum_{m=1}^{\infty}e^{-(k-2)m^2\beta^2/4\pi\tau_2}
    \frac{\abs{\eta(\tau)}^4}{\abs{\vartheta_{11}\bigl(-\frac{im\hat{\beta}}{2\pi},\tau\bigr)}^2}.
\end{multline}
The Hagedorn behavior of the partition function is manifest in the
limit $\tau_2\rightarrow 0$ (in field theory this is a UV limit.\footnote{As always, this is related to an IR divergence due to a tachyonic mode by a modular transformation.}) In this limit the
partition function diverges exponentially for $\beta<\beta_H$, which
corresponds to a Hagedorn transition at $T=T_H$. There is another
exponential divergence for $\tau \rightarrow \infty$ due to the
usual bosonic String tachyon, but it has no temperature dependence.
Poles at zeros of the theta-function are interpreted in
\cite{Maldacena:2000kv} as volume divergence due to worldsheet
instantons which can make their way to the boundary.

The calculation of the leading order behavior in the
$\tau_2\rightarrow 0$ limit is independent of the value of $\tau_1$
which we set to zero. We also set $\mu=0$ which will be the focus of
this paper. The contribution of the internal CFT ($\mathcal{M}$) is
\begin{equation}
    \sum_{h,\bar{h}}D(h,\bar{h})q^hq^{\bar{h}}\sim e^{\frac{\pi c_{int}}{6\tau_2}},
\end{equation}
combined with the contribution for the eta-functions, theta-function and exponential we find the integrand of
\eqref{partition-strip} behaves as
\begin{gather}
    \sim\exp\left[-\frac{k\beta^2}{4\pi\tau_2}+\frac{\pi\bigl(24-\frac{6}{k-2}\bigr)}{6\tau_2}\right].
\end{gather}
Thus, the integrand is exponentially divergent for $\beta<\beta_H$ where,
\begin{equation}\label{hagedorn_temp}
    \beta^2_H=\frac{4\pi^2}{k}\bigl(4-\frac{1}{k-2}\bigr),
\end{equation}
which signals the Hagedorn behavior of the theory. A quick check of the result is to consider the flat space
limit ($k\rightarrow\infty$) which produces as expected the result of \cite{Atick:1988si},
\begin{align*}
    \beta_{H}^{({\rm flat})} = \lim_{k\rightarrow\infty}k\beta_{H}^2 = 16\pi^2.
\end{align*}
The extra factor of $k$ is a necessary part in the double scaling limit which gives the flat-space geometry
correctly.

As discussed in \cite{Atick:1988si}, the Hagedorn behavior can also be seen as the appearance of the
Atick-Witten tachyon \footnote{The best definition for a tachyon in Euclidean theory is an exponential
divergence in the partition function related to IR physics in space time.} in the action of a string wrapping
the compact time coordinate as we raise the temperature (shrink the time circle). A similar calculation in
thermal \ads3, where one wraps the non contractible time circle with a winding string, reproduces the same
result as the one loop partition function. We shall return to this calculation in the sequel.

The interpretation of the Hagedorn behavior in \ads{} space was clarified in \cite{Barbon:2001di,Barbon:2002nw,Barbon:2004dd}. Starting from
low temperature the stable and dominant phase is a thermal \ads{} which can be viewed as a gas of strings at
thermal-equilibrium. As we raise the temperature we cross the Hawking-Page temperature where thermal \ads{} is no
longer the dominant phase but it is meta-stable. If we continue to raise the temperature forcing the system to
stay in the thermal \ads{} state (i.e an overheated \ads{}) we are bound to reach the Hagedorn temperature where the
thermal \ads{} system develops instability in perturbations theory and the system flows to another phase. For \ads3
with vanishing chemical potential the only other dominant absolutely stable phase is the Euclidean $BTZ$ black
hole (with no angular momentum nor any RR charge), thus, as we cross the Hagedorn temperature we expect that the
Thermal \ads3 must collapse to the Euclidean $BTZ$ black hole.

\begin{figure}[ht]
  \begin{center}
    \includegraphics[angle=270,width=0.9\textwidth]{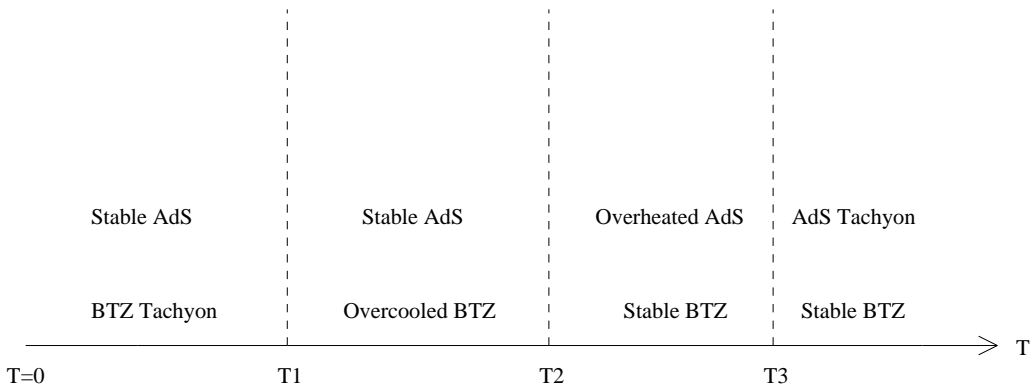}
    \caption{Phase diagram for the asymptotic \ads3 boundary conditions with chemical potential $\mu=0$. As we explained, the $EBTZ$ tachyon is winding around the angle of the $EBTZ$ while the \tads3 tachyon winds around the compact time of the \tads3. The various temperatures are: $T1=\frac{1}{2\pi\sqrt{k}}\left(4-\frac{1}{k-2}\right)^{1/2}$ , $T2=\frac{1}{2\pi}$ , $T3=\frac{\sqrt{k}}{2\pi}\left(4-\frac1 {k-2}\right)^{-1/2}$.}
    \label{fig:phase-diagram}
  \end{center}
\end{figure}

A very interesting property, unique to \ads3, is the diffeomorphism between \tads3 with identification $\tau$
and $EBTZ$ with identification $\frac{-1}{\tau}$, setting $\mu=0$ the diffeomorphism is between \tads3 at
temperature $T$ and $EBTZ$ at temperature $\frac{1}{4\pi^2T}$. Therefore we can move backwards on the phase diagram just described - starting with $EBTZ$ at high temperature cooling the system through the Hawking-Page temperature to a overcooled (meta-stable) $EBTZ$ up to the point we reach yet another Hagedorn temperature where the system must collapse to \tads3. The complete phase diagram is sketched in figure \ref{fig:phase-diagram}.
%\begin{center}
%\EPSFIGURE{phase.eps, angle=270,width=0.9\textwidth}{\label{fig:phase-diagram}Phase diagram for the asymptotic \ads3 boundary conditions with chemical potential $\mu=0$. As we explained, the $EBTZ$ tachyon is winding around the angle of the $EBTZ$ while the \tads3 tachyon winds around the compact time of the \tads3. The various temperatures are  : $T1=\frac{1}{2\pi\sqrt{k}}\left(4-\frac{1}{k-2}\right)^{1/2}$ , $T2=\frac{1}{2\pi}$ , $T3=\frac{\sqrt{k}}{2\pi}\left(4-\frac1 {k-2}\right)^{-1/2}$.}
%\end{center}

From the $EBTZ$ point of view there is also an analog of the tachyon winding the time circle in \tads3. Using the diffeomorphism between $EBTZ$ and \tads3 we learn that when we lower the $EBTZ$ temperature a string winding the angular coordinate develops a tachyonic mode exactly where the diffeomorphic \tads3 develops the Atick-Witten tachyon,
\begin{equation}
    T_{c}^{({EBTZ})} = \frac{1}{2\pi\sqrt{k}}\left(4-\frac{1}{k-2}\right)^{1/2}
    =\frac{4\pi^2}{T_H^{({TAdS})}}.
\end{equation}
Note that the proper size of the angular circle of the $EBTZ$ black
hole when one reaches this temperature is of the order of the String
scale. In spite of the fact we have obtained this result from
Euclidean considerations, it has an interesting Lorentzian
interpretation as well. One can continue the $EBTZ$ black hole back
to Lorentzian signature by rotating the time circle. Near the point
where the mode becomes tachyonic, the resulting black hole is small
with horizon area of the order of the String scale. Thus, we have
calculated exactly the point where the tachyon which is expected to
be generically behind the horizon of the $BTZ$ black hole \footnote{Due to the orbifold singularity behind the horizon.} "leaks" outside the horizon. This is closely related to the
phenomenon discussed (mainly in the \ads{5} context) in
\cite{Horowitz:2006mr}.

The above result is exact in $\alpha'$ (equivalently in $k$)
enabling to probe the theory in the strongly curved limit. An
immediate feature of the above formula, is the behavior for $k=3$
where,
\begin{equation}
    k\rightarrow 3:\qquad T_{c}^{({EBTZ})}=T_{H}^{({TAdS})}=T_{HP}=2\pi.
\end{equation}
Consequently, the region $k<3$ does not make any sense physically \footnote{Naively, one expects the theory to break down at $k=2$ where the energy momentum tensor ceases to exist but not before.} because the theory develops an instability before one reaches the phase transition point. In fact, as explained in
\cite{Giveon:2005mi}, when $k=3$ the Lorentzian $\SL{2,\Rf}$ model
loses its vacuum. Therefore the interpretation of the model as
string propagating in \ads3 is no longer valid. It will be nice to
understand better the relation between these two pathologies, we
further discuss this in the next subsection.

One can analyze the properties of the Atick-Witten tachyon \footnote{We thank S. Minwalla for very instructive discussions on this point, and in particular on the delocalization of the tachyon.} by a minisuperspace calculation of a string winding around the time circle. The details of the calculation are somewhat
technical so we leave it to appendix \ref{z-tit-min}. The main
result of the calculation is the eigenvalue equation describing the
mode \eqref{mini-eigenvalue},
\begin{equation}
    E\Psi(\rho)=\frac{1}{2k}\left[-\partial_\rho^2-2\coth(2\rho)\partial_\rho
    +\left(\frac{\beta k}{2\pi}\right)^2-a\right]\Psi(\rho),
\end{equation}
where $a$ is related to the zero-point energy constant. It is calculated, by employing canonical quantization,
in appendix \ref{z-tit-min} as well. The winding string moves in a potential which is bounded at infinity. While
naively from the metric one expects an infinitely high potential as $\rho\rightarrow\infty$, the presence of the
B-field in the background cancels the $e^{2\rho}$ term coming from the geometry. At large $\rho$ the wave
function of the winding mode has the following dependance (see \eqref{mini-wf})
\begin{align}
    \Psi_j(\rho) \sim e^{+2j\rho}
    &&
    j = -\frac12+is
    &&
    E_j=\frac1{2k}\left(-4j(j+1)+\frac{\beta^2k^2}{4\pi^2}+a\right),
\end{align}
where $s$ is a real parameter. After fixing the constant $a$, one
discovers that at the Hagedorn temperature $\beta=\beta_H$ the
lowest mode ($s=0$) has exactly zero energy $E_{-1/2}=0$. It is
important to note that the winding mode is delta-function
normalizable.\footnote{In the Klein-Gordon norm there is a volume
factor $\sim \sinh(2\rho)$ that cancels the exponential decay of the
wave function.} This indicates that the condensation of this mode is
not a strictly local process. As we discussed in
\S\ref{y-tit-rev-sing} this oddity is related to the fact that the
spacetime CFT is singular, and it is removed when the background is
appropriately regulated.

\subsection{More on $k=3$}

As argued above, the ensemble does not have sensible thermodynamic properties below $k=3$. This fits well with
the fact that the Lorentzian vacuum of $\SL{2,\Rf}$ does not exist below $k=3$, but one can be more precise. A
host of special effects which take place at $k=3$ were discussed in \cite{Giveon:2005mi}.\footnote{These effects
are associated to the black hole/black string transition which occurs exactly at $k=3$. } In addition to
projecting out the vacuum, it was found that the $BTZ$ black hole becomes non normalizable and that the correct
description below $k=3$ is by weakly coupled long strings (one can also check that the gap for long strings
disappears when $k=3$).

\begin{figure}[ht]
  \begin{center}
    \mbox{
      \subfigure[$k>3$\label{pot1}]{\scalebox{0.3}{\includegraphics[width=1\textwidth]{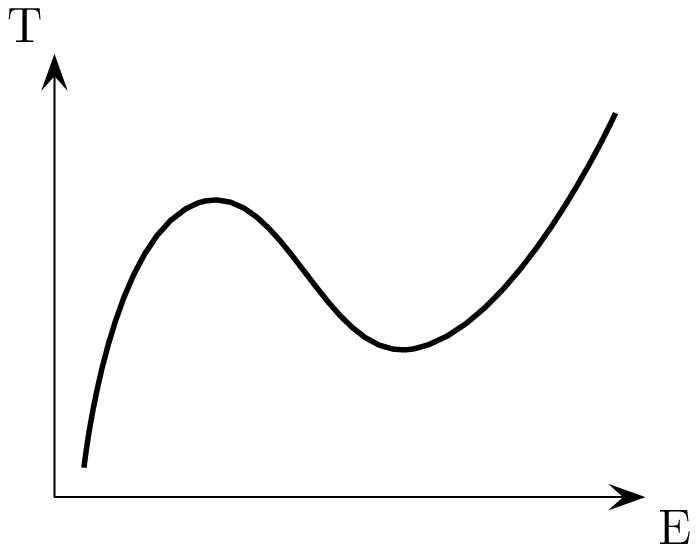}}} \quad
      \subfigure[$k=3$\label{pot2}]{\scalebox{0.3}{\includegraphics[width=1\textwidth]{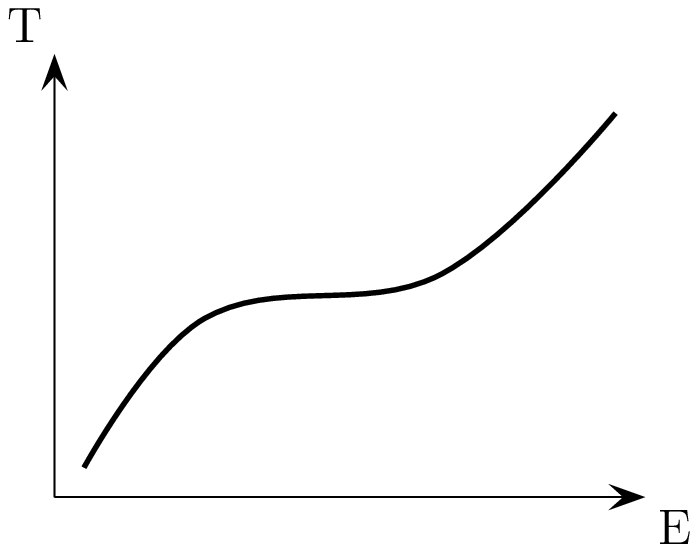}}} \quad
      \subfigure[$k<3$\label{pot3}]{\scalebox{0.3}{\includegraphics[width=1\textwidth]{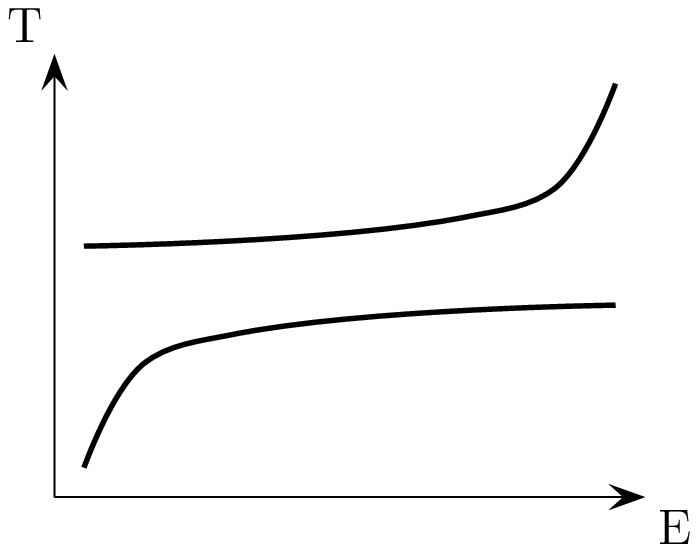}}} \quad
      }
    \caption{The vertical axis is the temperature of the system and the horizontal axis is the typical energy of an excitation. We exhibit the transition in the form of the phase diagram from $k>3$ to $k=3$ and finally to $k<3$. The most dramatic change is the appearance of a limiting temperature when $k<3$, which is closely related to the fact that typical high energy excitations are not black holes anymore but long fundamental strings.}
  \end{center}
\end{figure}

This culmination of observations makes the connection to our result
more obvious.\footnote{We are grateful to O.~Aharony for discussions
on this point.} For $k>3$, the thermodynamic description is the
standard one, as depicted in figure \ref{pot1}. A generic high
energy state is the black hole, and the gas of particles ceases to
exist beyond the Hagedorn temperature. When $k=3$, the states
smoothly go over to each other, and it is manifest that the Hagedorn
and Hawking-Page temperatures match. This situation is depicted in
figure \ref{pot2}. Lastly, the only thing that can happen below
$k=3$ and be consistent with everything said above is that the $BTZ$
black hole can never be reached (the system has a physical limiting
temperature), and a generic state is just a weakly coupled long
string. This is displayed in figure \ref{pot3} where the line of
putative black holes, above a gap in temperature, is also exhibited
(of course, it can never be reached) for making it easy to visualize
how the transition \footnote{From a system that has no limiting
temperature to one that has.} at $k=3$ occurs.

\section{The competing winding modes picture}\label{z-tit-comp}
\tads3 and $EBTZ$ are different ways of extending the $S^1\times S^1$ boundary to a 3 dimensional manifold. In the
two cases, one fills different $S^1$'s into a disk $D_2$. When one of the backgrounds becomes unstable, an
Atick-Witten tachyon appears which is a winding mode on the non-contractible circle. Hence, the flow dynamics is
controlled by a competition between a geometric capping, by $D_2$, and a Stringy capping, by a winding tachyon.
Fortunately, as we will discuss in this section, in String theory these two ways of capping are
indistinguishable (in a very precise technical sense). This becomes apparent in a dual formulation which
contains a sine-Liouville (s-L) theory, and is the main motivation for investigating this relation.

The picture we will end up with consists of two deformations of
linear dilaton by s-L walls, whose profile grows as we go into the
bulk of space. They both tend to shrink their respective circles.
The one with the larger coefficient on the worldsheet "wins" and
caps its circle first. The problem of quantifying the phase
transition is now just a problem of comparing the coefficients of
two operators on the worldsheet. Further, we show that this
perturbation of s-L becomes the Atick-Witten tachyon vertex operator
exactly at the Hagedorn temperature. Consequently, the issues of
geometric capping and tachyon condensation are closely related in
our backgrounds.

This approach also provides information about the CFT that corresponds to the unstable fixed point "between" the $EBTZ$ and \tads3 (the sense in which this is "between" the two phases will be specified precisely). This is discussed in \S\ref{z-tit-mid}.

\begin{figure}[ht]
  \begin{center}
    \mbox{
      \subfigure[Thermal \ads3\label{adsfig}]{\scalebox{0.33}{\includegraphics[width=1\textwidth]{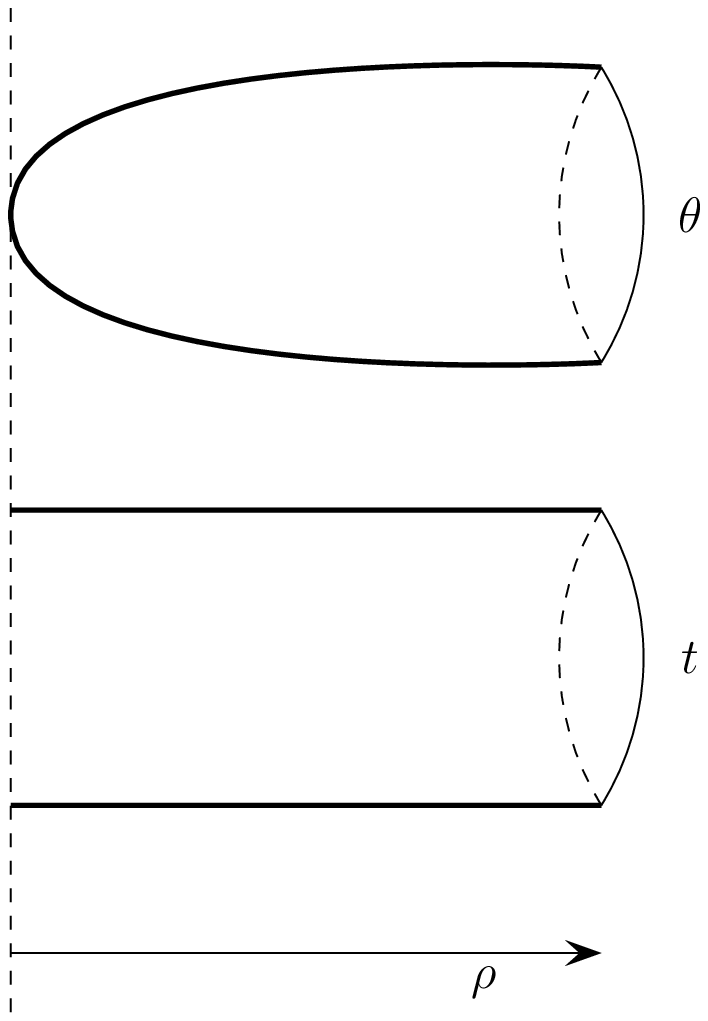}}} \qquad
      \subfigure[Euclidean $BTZ$\label{btzfig}]{\scalebox{0.33}{\includegraphics[width=1\textwidth]{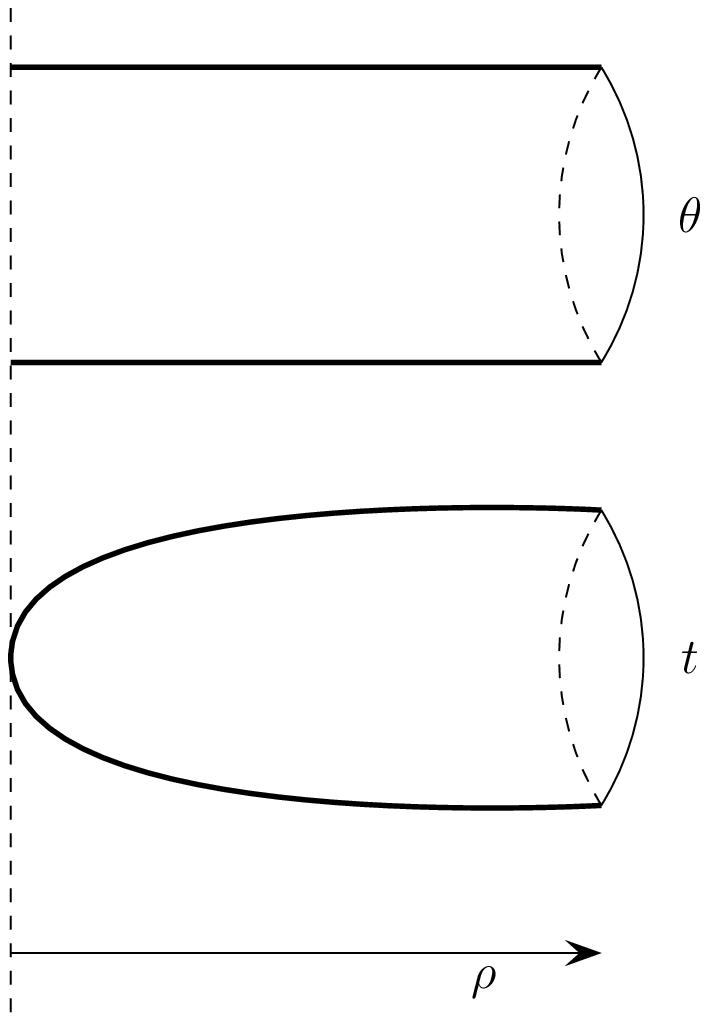}}}
      }
    \caption{The topological structure of thermal \ads3 on the left and Euclidean $BTZ$ black hole on the right. The non contractible circles are the thermal (time) circle for \tads3 and the angular (spatial) circle in $EBTZ$.}
  \end{center}
\end{figure}

Intuitively, the idea is as follows. Thermal \ads3, in which the disk fills up the angular direction, is represented in figure \ref{adsfig}. Suppose the temperature is high enough so that a tachyon develops on the non contractible circle. On general grounds, one would expect the thermal circle to pinch and the space to cut and begin from some larger value of the radial coordinate $\rho\geq\rho_0$. The resulting space is easy to visualize. It has a contractible thermal circle and a non contractible angular variable. This is precisely what we need for a Euclidean black hole. Indeed, the Euclidean $BTZ$ black hole (whose topological structure is exhibited in figure \ref{btzfig}) is our candidate for the end point of tachyon condensation.

The picture is reversed when we consider a very cold Euclidean $BTZ$ black hole- the disk fills the Euclidean time
circle and the tachyon develops along the angular circle contracting it. Therefore, tachyon condensation
switches the topology to that of a \tads3.

However, we will see that the dynamics is not so simple. As in the $\Cf/\Zf_N$ case considered in \cite{Adams:2001sv}, there is a need for an outgoing shell to change some boundary conditions at infinity (after infinite RG flow time). Otherwise, the flow can not end in a smooth background. The physics we find here is very similar in some respects.

\medskip
For convenience we outline the procedure we employ to derive our dual description
\begin{itemize}
\item Begin with either \tads3 of $EBTZ$ (the end result will be the same).
\item By twice T dualizing non contractible circles, we write the background as
    $\frac{\SL{2,\Rf}}{\U{1}}\times\U1$. This is done in \S \ref{y-tit-comp-tads}.
\item Using FZZ duality (reviewed in \S \ref{y-tit-comp-fzz}), the latter is a linear dilaton theory
    perturbed by a winding mode condensate, a ''wall", in the bulk.
    The geometric cap is replaced by a winding mode cap.
\item
The model is now a linear dilaton with two circles and a winding
mode condensate on one of them. If one is at the Hagedorn
temperature, turning on the Atick-Witten tachyon on the Euclidean
time direction amounts to turning on another sine-Liouville
interaction on the other circle. The picture is now completely
symmetric between the spatial and time circles where there are two
vertex operators of the same kind.
\end{itemize}
In the range of temperatures in which both \tads3 and $EBTZ$ are
perturbatively stable, the thermodynamic description (see figure
\ref{pot1}) suggests that there is an unstable fixed point between
these two stable fixed points. One can guess how to construct this
unstable point.\footnote{Of course, the fact it is unstable at
intermediate temperatures implies that it should be, in some sense,
a singular solution. This is analogous to the small black hole in
the \ads{5} ensemble.} What we have basically done is to describe
two distinct ways to cap, in a non singular way, the model $linear\
dilaton\times S^1\times S^1$. In \tads3 we have filled with a disc
one of the circles while in $EBTZ$ we have filled with a disc the
other. To obtain the intermediate point one needs to treat the two
circle more symmetrically. One can either
\begin{itemize}
\item
Not cap either of the circles. In this case the model is simply
$linear-dilaton\times \mathbb{T}^2$. This point does not exist in
the space of perturbative String theory. It may exist once
non-perturbative effects are taken into account (similar to the
distinction between LST and DLST backgrounds
\cite{Berkooz:1997cq,Seiberg:1997zk,Giveon:1999px,Aharony:1998ub}).
\item
Cap both circles at the same time. If one can truncate the dynamics
to the coefficient of the two sine-Liouville interaction then such a
point must exist simply because there must exist an unstable fixed
point between two stable fixed point if there in one real parameter.\footnote{We will see that the situation is slightly more complicated than that.} This capping is perturbative in $g_s$.
\item Some other way of capping, such as a Liouville cap (without winding condensation).
\end{itemize}
We return to the question of constructing the unstable phase in \S \ref{z-tit-mid}. In \S \ref{y-tit-comp-reg}, we discuss a way to regulate the singularities reviewed in \S\ref{y-tit-rev-sing} and explain the physical consequences of such a regulator.

\subsection{A review of the FZZ duality}\label{y-tit-comp-fzz}

The FZZ duality, first conjectured by V.Fateev, A.Zamolodchikov and
Al.Zamolodchikov \cite{FZZ}, is a correspondence  between the
~$\SL{2}_k/\U1$ coset CFT and sine-Liouville theory (see
\cite{Kazakov:2000pm,Aharony:2004xn,Kim:2005av,Fukuda:2001jd} and
references therein for relevant works on the subject). We shortly
review the duality following \cite{Kazakov:2000pm}.

The $\SL{2}_k/\U1$ coset CFT
\cite{Witten:1991yr,Mandal:1991tz,Elitzur:1991cb} also known as the
2-dim Euclidean black hole or cigar model, describes the geometry
\begin{align}\label{cigar-2d}
    ds^2 = k\left(dr^2+\tanh^2r\, d\theta^2\right)
    &&
    \Phi-\Phi_0=-2\log\cosh r,
\end{align}
where $\theta$ is a periodic coordinate $\theta\cong\theta+2\pi$ and
$r$ ranges from 0 to $\infty$. At the boundary of space,
$r\rightarrow\infty$, the model reduces to a product of a linear
dilaton and a circle, and at the tip of the cigar, $r=0$, the circle
pinches. The geometry is smooth everywhere, and the String coupling
$g_s=e^{\Phi_0}$ at the tip is finite. The level $k$ is a free
parameter which governs the size of the cigar. The mass of the
underlying Lorentzian black hole is related to the String coupling
at the tip of the cigar $M\propto e^{-2\Phi_0}$.

The central charge of the cigar CFT \eqref{cigar-2d} is,
\begin{gather}\label{cigar-central}
    c = \frac{3k}{k-2}-1.
\end{gather}
For the special case $k=\frac94$ the solution is a two dimension bosonic String theory dual to a $c=1$ matrix
model. For larger values of $k$ one needs to multiply the model by a suitable CFT to complete the central charge
to 26. An important set of observables corresponds to momentum and winding modes on the cigar $V_{j,m\bar m}$.
These have scaling dimensions
\begin{align}
    \Delta_{j,m,\bar m} = -\frac{j(j+1)}{k-2}+\frac{m^2}{k}
    &&
    \bar \Delta_{j,m,\bar m} = -\frac{j(j+1)}{k-2}+\frac{\bar m^2}{k}.
\end{align}
The parameters $m,\bar m$ are related to the winding and momentum quantum numbers around the $\theta$ circle at
large $r$,
\begin{align}
    m = \frac{n+w k}2
    &&
    \bar m = -\frac{n-w k}2
    &&
    n,w\in\Zf.
\end{align}
The geometry of the cigar, which is topologically a plane, makes it obvious that the cigar background conserves
only the momentum $m-\bar m$ and not the winding $m+\bar m$,

String perturbation theory (genus expansion) is an expansion in $e^{2\Phi}$, which can be thought as an
expansion in the String coupling at the tip $e^{2\Phi_0}\sim\frac1M$. The geometry has radius of curvature k,
and thus the worldsheet is weakly coupled for large $k$. However the coset model is well defined for small
values of $k$ as well.\footnote{There are some $\frac1k$ corrections to the action, as explained in
\cite{Dijkgraaf:1991ba}.}

The dual sine-Liouville theory is defined by the 2-dim Lagrangian density
\begin{equation}\label{SL-lagran-2d}
    4\pi \mathcal{L} = (\pd \phi)^2+(\pd x)^2+Q\hat R\phi+\lambda e^{b\phi}\cos R\left(x_L-x_R\right),
\end{equation}
where the $x$ direction is periodic with
\begin{align}
    x\cong x+2\pi R
    &&
    R = \sqrt k,
\end{align}
and the theory has a linear dilaton  $\Phi-\Phi_0 =-\frac{Q}2\phi$. The central charge of this theory is
$c=2+6Q^2$. Comparing with \eqref{cigar-central} we set,
\begin{equation}
    Q^2=\frac1{k-2}.
\end{equation}
The s-L interaction is the lowest lying winding mode (winding number equals one) around the circle $x$, the
exponent $b$ is fixed by requiring that the interaction is marginal,
\begin{equation}\label{SL-exp-b}
    \frac1{4}R^2-\frac14b(b+2Q) = 1 \rar1 b = -\frac1{Q}=-\sqrt{k-2}.
\end{equation}

There are two inequivalent models: $\lambda\not=0$ and $\lambda=0$. In the first case, the coefficient
$\lambda>0$ can be rescaled to any other positive value by a shift in $\phi$ (and re-absorbed in the String
coupling). If we set $\lambda$ to zero we are left with the linear dilaton theory (with the coupling diverging
at $\phi\rightarrow-\infty$) which is ill defined in perturbation theory. In \cite{Kazakov:2000pm} the authors
treated the s-L theory as a perturbation of Liouville theory, i.e adding a Liouville potential
$\mu_L\phi\,e^{-2\phi}$ and treating $\lambda$ as a perturbation. Varying the radius $R$ of the $x$ circle they
found that at a critical of radius $R=\sqrt k$ one can take the limit $\mu_L\rightarrow 0$ with no singularities. \footnote{The calculation in \cite{Kazakov:2000pm} was done only for $k=9/4$ due to the technical need of using matrix model results. The fact there is an effective wall due to the s-L interaction is well established for any
$k$ by investigating three point functions (for more details see \cite{Aharony:2004xn} and references therein).}
Naively, the s-L theory has an infinite coupling region ($\phi\rightarrow-\infty$), but the interaction terms
generates a wall preventing particles from reaching the strong coupling area (similar to the cosmological
constant potential $\mu_L\phi e^{-2\phi}$ in the more familiar Liouville theory).

Wave functions of s-L theory behave at large $\phi$ as
\begin{equation}
    \Psi(\phi) \sim e^{\left(Q-\frac1{Q}\right)\phi}.
\end{equation}
At large $k$ the wave function $\Psi$ goes rapidly to zero at the
weak coupling area ($\phi\rightarrow\infty$), thus, the theory is
effectively strongly coupled. On the other hand if $k\rightarrow 2$
($Q\rightarrow \infty$) the wave function $\Psi$ is supported at
large $\phi$ away from the potential wall. Consequently, s-L theory
is strongly coupled when the cigar CFT becomes weakly curved and
vice-versa.

The observables $V_{j,m,\bar m}$ of the cigar are mapped to vertex
operators of s-L which have the following large $\phi$ behavior (for
large k this is simply the requirement that the asymptotics of the
vertex is the same):
\begin{equation}
    V_{j,m,\bar m}\leftrightarrow e^{ip_Lx_L+ip_Rx_R+\beta\phi},
\end{equation}
with
\begin{align*}
    p_L = \frac{n}{R}+w R &&
    p_R = \frac{n}{R}-w R &&
    \beta = 2 Q j.
\end{align*}
At large $\phi$ and $r$ ($\phi,r\rightarrow\infty$) both s-L theory and the cigar model describe a cylinder with
linear dilaton implying the identification (at large k)
\begin{align}
    r\sim -Q \phi
    &&
    \theta\sim \frac{x}{\sqrt k}.
\end{align}
FZZ duality is the statement that the cigar coset model and s-L theory are exactly equivalent as conformal field
theories. As explained above, this is interpreted as a strong-weak duality of the worldsheet theories.

S-L correlation functions have a KPZ behavior \cite{Knizhnik:1988ak}
implying that the partition sum has the following genus expansion,
\begin{equation}\label{kpzscl}
    \mathcal{F}(\lambda,g_s)=\sum_{h=0}^{\infty}\mathcal{F}_h\left(g_s\lambda^{-\frac1{k-2}}\right)^{2(h-1)}.
\end{equation}
Thus, $g_s\lambda^{-\frac1{k-2}}$ is an effective String coupling and one can set $g_s=1$ for convenience. On the other hand, genus expansion of String theory on the cigar is related to the String coupling at the tip of the cigar $g_s^2\sim 1/M$ (where M is measured in planck units). Therefore in the FZZ duality the correspondence
of the genus expansions suggests,
\begin{equation}
    M\leftrightarrow \lambda^{\frac{2}{k-2}}.
\end{equation}

There is no complete proof of the bosonic conjecture (see \cite{Giribet:2007uh} for recent review), but strong evidence comes from comparison of 2-pt and 3-pt
functions. The supersymmetric version of the duality is given in \cite{Giveon:1999px}, Hori and Kapustin \cite{Hori:2001ax} used gauged linear sigma-model techniques to show that in this case the duality is an example of mirror symmetry. For recent discussion on the supersymmetric duality and it's relation to the bosonic conjecture see \cite{Giveon:2003wn,Israel:2004jt,Maldacena:2005hi}

\subsection{\tads3 and $EBTZ$ using FZZ duality}\label{y-tit-comp-tads}

\subsubsection{Step 1: The background as $cigar\times S^1$}\label{x-tit-comp-back}

Recall the solution of the thermal \ads3 background \footnote{We use the wedge product normalization ~$a\wedge b
= \frac12(a\otimes b +b\otimes a)$.}
\begin{align}\label{TADS-background}
    &ds^2=k\left(d\rho^2+\cosh^2\rho\, dt^2+\sinh^2\rho\, d\theta^2\right)+ds^2_{\perp}
    \cr
    &B_{(2)}=-2ik\sinh^2\rho\, dt\wedge d\theta
    \cr
    &e^{\Phi}=g_s,
\end{align}
with the identifications
\begin{equation}
    \theta\cong\theta+2\pi\csp2
    t\cong t+\beta.
\end{equation}
The B-field is imaginary since it is an analytic continuation of a
real B-field in Lorentzian space. We T-dualize \cite{Giveon:1994fu}
the non contractible time circle to obtain
\begin{align}\label{first-t-dual}
    &ds^2=k\left[d\rho^2+\tanh^2\rho(\,d\theta-i\,d\tilde t)^2+d{\tilde t}^2\right]
    +ds^2_\perp
    \cr
    &B_{(2)}=0
    \cr
    &\Phi'=\log(g_s)-\log\left(\cosh^2\rho\right)-\log\left(\frac{k\beta^2}{4\pi^2}\right),
\end{align}
with the identifications,
\begin{equation}
    \theta\cong\theta+2\pi\csp2
    \tilde t\cong \tilde t+\frac{4\pi^2}{k\beta}.
\end{equation}

Note that some key features in the geometry at infinity have apparently changed. In \eqref{TADS-background}, the
boundary is only conformal to $\mathbb{T}^2$ with a scale factor that diverges at infinity. In
\eqref{first-t-dual} the metric of the boundary, in the String frame, is a finite $\mathbb{T}^2$. The
interpretation of this result is that in the original picture, \eqref{TADS-background}, the energy of winding
strings that go to the boundary is a finite constant (this is derived in a mini-superspace framework in appendix
\ref{z-tit-min}.).

The background in the $\rho$-$\theta$ plane is the familiar
Euclidean cigar background (also known as the 2-dim Euclidean  black
hole) studied in \cite{Witten:1991yr,Mandal:1991tz,Elitzur:1991cb},
if we define $d\chi=\,d\theta -i\,d\tilde t$ and treat it as if it
has the usual reality condition of a real scalar.\footnote{One can
write down the dictionary for vertex operators, taking this factor
$i$ into account. However, we will not need to deal with it for a
reason which will soon become clear.} We can then use known results
about the cigar. For example, as a quick check, one can compare the
central charge of the $cigar\times S^1$ background with the central
charge of \tads3,
\begin{align*}
    c_{({AdS}_3)} =\frac{3k}{k-2}= 3+\frac{6}{k-2}
    &&
    c_{({cigar}\times S^1)} = \left(2+\frac{6}{k-2}\right) + 1.
\end{align*}
This is, of course, not surprising.

Next, we apply another T-duality in the decoupled $\tilde t$ circle
direction (such that the $d\chi$ direction not effected by the
transformation), we find the following background (denoting the
T-dual coordinate to $\tilde t$ by $\varphi$)
\begin{align}\label{cigar-metric}
    &ds^2=k\left(d\rho^2+\tanh^2\rho\,d\chi^2+d\varphi^2\right)
    +ds^2_\perp
    \cr
    &B_{(2)}=0
    \cr
    &\Phi'=\log(g_s)-\log\left(\cosh^2\rho\right)\rar1 g_s'=g_s\bigm/\cosh^2\rho,
\end{align}
with the identifications,
\begin{equation}
    \chi\cong\chi+2\pi\csp2
    \varphi\cong \varphi+\beta.
\end{equation}
Our result can be summarized as follows
\begin{equation}\label{ciga}
    TAdS_3=\left(\frac{\SL{2,\Cf}_k}{SU(2)}\Bigm/U(1)\right)\times U(1)_{\beta,\,k},
\end{equation}
where the notation $U(1)_{\beta,\,k}$ is there to remind us that the
inverse temperature is $\beta$ but the proper size of the decoupled
circle is $\beta\sqrt k$. This decomposition is similar to the one
suggested in \cite{Horne:1991gn} for the Lorentzian signature
backgrounds. One immediate surprise is encountered if one tries to
repeat this exercise in the case of the $EBTZ$ black hole. The
easiest way to do it for the black hole is to exploit its
equivalence to a Thermal \ads{3} with a different temperature,
\begin{equation}\label{cigb}
    EBTZ=\left(\frac{\SL{2,\Cf}_k}{SU(2)}\Bigm/U(1)\right)\times U(1)_{{{4\pi^2}/\beta},\,k}.
\end{equation}

In both \eqref{ciga} and \eqref{cigb}, the asymptotic form of the
metric is just $linear-dilaton\times S^1\times S^1$. For both \tads3
and $EBTZ$ one pinches the circle of proper size $2\pi\sqrt k$ and
the other remains intact. {\it However}, the decoupled circles have
different proper size at infinity. This is surprising at first
sight, since one expected \tads3 and $EBTZ$ to be in the same
thermal ensemble (when they have the same temperature). Indeed, they
are in the same ensemble when one does pure gravity. String theory,
on the other hand, clearly has a non normalizable mode
distinguishing the two backgrounds. This is disguised in the
original description but it is manifest in this T dual frame, where
this non normalizable mode is a proper size of a free circle at
infinity.

Thus, it appears that \tads{3} with modular parameter $\tau$ and \tads{3} with modular parameter
$\frac{-1}{\tau}$ are not in the same thermal ensemble in the framework of String theory. One would like to have
an understanding of this pathology in the original frame as well. The naive way to characterize this is to
recall that long strings (in the original frame) have constant energy potential near the boundary. The value of
this constant is different in the two backgrounds as can be seen from the explicit computations in appendix
\ref{z-tit-min}. There is a more elegant description in the original frame which is explained in the sequel.

Consider the asymptotic form of \tads3, keeping the first sub-leading correction to the $B$ field
\begin{align}
    &ds^2\simeq k\left(d\rho^2+\frac{e^{2\rho}}{4}dt^2+\frac{e^{2\rho}}{4}d\theta^2\right)
    \cr
    &B_{(2)} = -2ik\left(\frac{e^{2\rho}}{4}-\Lambda_{(TAdS)}\right)\, dt\wedge d\theta.
\end{align}
where $\Lambda_{(TAdS)}=\frac12$. The identifications of \tads{3} with
modular parameters $\tau$ and $\frac{-1}{\tau}$ are respectively
\begin{align*}
    {TAdS}_3:&\quad  t\cong t+\beta~\qquad \theta\cong\theta+2\pi
    \cr
    {EBTZ}:&\quad  t\cong t+\frac{4\pi^2}{\beta}~\quad \theta\cong\theta+2\pi.
\end{align*}
We can bring the two conformal $\mathbb{T}^2$'s to have the same periodicities  by a shift in $\rho$, however
this shift changes the constant in the B-field. For concreteness shifting $\rho$ in $EBTZ$ results in
\begin{equation}
    \rho\rightarrow \rho+\log\frac{2\pi}{\beta}
    \rar1
    \Lambda_{(EBTZ)} - \Lambda_{(TAdS)} = \frac12\left(\frac{\beta^2}{4\pi^2}-1\right).
\end{equation}

One could think that this constant mode of the $B$ field is a an
unphysical pure gauge mode. This is not true in the thermal case
since the integral $$\int_{\mathbb{T}^2}B$$ plays the role of a
generalized Wilson line, which has a physical effect on the spectrum
of strings. Thus, the calculation above suggests that the invariant
way to characterize the non normalizable mode which distinguishes
the two backgrounds is indeed $\int_{\mathbb{T}^2}B$. This fits well
with the previous explanation involving the potential energy of long
strings, since it is exactly this Wilson line which sets this
constant.\footnote{Note that this mode exists only near the boundary
of space, since in the full background, of the circles of the two
torus in contractible.}

For completeness, it remains to show explicitly that this constant
translates directly to the volume of the two torus in the T dual
frame. To demonstrate this, we begin with the \tads{3} metric with
the undetermined constat $\Lambda$ in the B-field
\begin{equation}\label{dfrmdb}
    B_{(2)}'=-2ik(\sinh^2\rho+\delta\Lambda)\, dt\wedge d\theta.
\end{equation}
As before, we first T dualize the non contractible Euclidean time direction and get the following metric (with
$B=0$)
\begin{equation}\label{afteroneTduality}
    ds^2=k\left[d\rho^2+\frac{\left(d\tilde t-i\delta\Lambda\,d\theta\right)^2}{\cosh^2\rho}
    -2i\tanh^2{\rho}\, d\tilde td\theta
    +\tanh^2\rho(1-2\delta\Lambda)\, d\theta^2\right]+ds_\perp^2.
\end{equation}
We define new complex coordinates
\begin{align*}
    \chi = \sqrt{1-2\delta\Lambda}~\theta - i\frac{\tilde t}{\sqrt{1-2\delta\Lambda}}
    &&
    \tilde t' = \frac{\tilde t}{\sqrt{1-2\delta\Lambda}}.
\end{align*}
Repeating our procedure we apply a second T-duality on the new $\tilde t'$ to find the a deformed cigar background
\begin{align}\label{fresultoneTduality}
    &ds^2=k\left[d\rho^2+\frac{1-2\delta\Lambda}{(1-\delta\Lambda)^2-\delta\Lambda^2\tanh^2\rho}\left(\tanh^2\rho d\chi^2+d\varphi^2\right)\right]+ds_\perp^2 \cr
    &B_{(2)} = -2ik\frac{(1-\delta\Lambda)\delta\Lambda}{(1-\delta\Lambda)^2\cosh^2\rho-\delta\Lambda^2\sinh^2\rho}\,d\varphi\wedge d\chi \cr
    &\Phi' = \log(g_s) - \log\left(\cosh^2\rho\right)-\log\left(1+\frac{\delta\Lambda^2}{(1-2\delta\Lambda)\cosh^2\rho}\right)+\log(1-2\delta\Lambda)\cr
    &\chi\cong\chi +2\pi\sqrt{1-2\delta\Lambda}
    ~,\qquad
    \varphi\cong\varphi+\beta\sqrt{1-2\delta\Lambda}.
\end{align}
A fast consistency check is to verify that for $\delta\Lambda=0$ the
above result is the same as \eqref{cigar-metric}, i.e the usual
cigar metric.\footnote{The metrics we obtain with such T dualities
are always not real, but can be made real by analytically continuing
the Euclidean time. In particular, the background above, if
continued to Lorentzian signature is a perfectly well defined
solution to the equations of motion, and is worth understanding
better. It is interesting that one can generate non trivial
solutions in the T dual frame by dialing a seemingly trivial mode in
the original frame.}

To prove our assertion of the relation between the B field in the \ads{3} picture and the asymptotic volume in
the T dual frame we analyze the asymptotic form of (\ref{fresultoneTduality}), the B-field vanishes and the
asymptotic behavior of the metric is that of a $\mathbb{T}^2$ with radii
\begin{align*}
    R_\chi = \sqrt k\,\sqrt{1-2\delta\Lambda}
    &&
    R_{\varphi}=\sqrt k\,\frac{\beta}{2\pi}\,\sqrt{1-2\delta\Lambda},
\end{align*}
which is the original torus rescaled by $\sqrt{1-2\delta\Lambda}$~, as claimed.

Consequently, the Euclidean $BTZ$ black hole and \tads{3} have different asymptotic values of the generalized
Wilson lines, which inevitably leads to a non normalizable mode distinguishing them. This prevents any possible
RG flow which takes finite RG time from the usual black hole and \tads{3}. However, there can still exist flows
which emit a propagating shell affecting some boundary conditions as in \cite{Adams:2001sv}. Note that if we are
exactly at the Hawking-Page temperature, then this difference does not exist.

The implication of our result to Lorentzian physics requires
clarification. The integral $\int_{\mathbb{T}^2} B$ has no analogue
in the Lorentzian case since the time coordinate is non compact. In
particular, there is no obstruction for creating in a hot Lorentzian
\ads{3} an excitation which is the $BTZ$ black hole. The correct
interpretation of our result is that the thermal ensemble of this
system is somewhat pathological, and should be considered with care.
The micro-canonical ensemble, on the other hand, follows the
standard expectations.

This brings us to the need of establishing deformations of the model
which have sensible canonical thermodynamic descriptions. It is in
these cases where everything one expects from this system is
satisfied without subtleties. We discuss such a deformation in \S
\ref{y-tit-comp-reg}.

\subsubsection{Step 2: FZZ duality on the cigar}\label{x-tit-comp-SL}

So far we have written the background as $cigar\times S^1$. We still
have two kinds of caps - one is geometric (the cigar pinching) and
the other via a condensing winding mode on the separate $S^1$. In
String theory, however, these two kinds of caps are the same. The
cleanest example is the FZZ duality (which we reviewed before).

Applying the FZZ duality to the cigar metric, we find the s-L lagrangian (with additional circle),
\begin{equation}\label{SL-lagran}
    4\pi \mathcal{L} = (\pd \phi)^2+(\pd x)^2+(\pd\varphi)^2+Q\hat R\phi+\lambda e^{b\phi}\cos
    R_x\left(x_L-x_R\right).
\end{equation}
The central charge of this theory is $c=3+6Q^2$, the dilaton slope is
\begin{align}
    \Phi-\Phi_0 =-\frac{Q}2\phi
    &&
    Q^2=\frac1{k-2}.
\end{align}
The radii of the circles $x$ and $\varphi$ are
\begin{align}
    &x\cong x+2\pi R_x
    &&    R_x = \sqrt k &
    \cr
    &\varphi\cong \varphi+2\pi R_\varphi
    &&
    R_{\varphi}=\frac{\beta\sqrt k}{2\pi} \ ,&
\end{align}
and the exponent $b$ calculated in \eqref{SL-exp-b}. We adopt the view of \cite{Kazakov:2000pm} defining the s-L
theory as the $\mu_L\rightarrow0$ limit of a Liouville theory with a s-L interaction.

The s-L interaction term in \eqref{SL-lagran} is clearly a winding
mode on the $x$ circle. Tracing back the FZZ and T dualities, the
winding around the non contractible time circle (which the
Atick-Witten tachyon of \tads3 winds around above the Hagedorn
temperature) is related to  winding  around the $\varphi$ circle in
the s-L theory.\footnote{The mapping of these states is exact and
does not involve subtleties in the mapping of the corresponding
vertex operators.}

The picture now is more symmetric - both circles can be capped with winding mode condensation. Roughly, one can
consider the more general deformation of the two circles by
\begin{equation}\label{two-defor}
{\cal L}={\cal L}_0+
    \lambda_1e^{b_1\phi}\cos R_x(x_L-x_R)~+~
    \lambda_2e^{b_2\phi}\cos R_\varphi(\varphi_L-\varphi_R),
\end{equation}
where ${\cal L}_0$ is the $linear-dilaton\times \mathbb{T}^2$ lagrangian, and the values of $b_1$ and $b_2$ are
determined by the marginality requirement (at tree level)
\begin{align}
    \frac14R_x^2 - \frac14 b_1(b_1+2Q) = 1 &&
    \frac14R_\varphi^2 - \frac14 b_2(b_2+2Q) = 1~.
\end{align}
We can shift $\phi$ to change both coefficients $\lambda_1$ and
$\lambda_2$ and only the ratio $\eta=\lambda_1^{1/b_1}/
\lambda_2^{1/b_2}$ has physical significance. Allowing the two
deformation theory \eqref{two-defor} to flow to its IR fixed point
we expect that, for generic value of $\eta$, one of the deformations
dominates over the other, driving the other to zero (and shrinking
the $S^1$ around which the dominant perturbation winds). Simply, the
deformation whose associated wall is closer to the weakly coupled
boundary is expected to dominate.

The analysis whether such a flow generates propagating waves which can, after infinite RG time flow, change the
boundary conditions (namely the proper size of the circles) is left for future work. It would also be
interesting to study \eqref{two-defor} using perturbations around Liouville but we leave this for future work as
well. A more careful account of the possible flows is postponed to \S\ref{z-tit-mid}.

A special point which does not suffer from the difficulties of
changing the boundary conditions is the Hawking Page point where all
the circles are of the same size. The physical parameter is
$\eta_s=\lambda_1/\lambda_2$, and the theory has enhanced symmetry
at $\eta_s=0,1,\infty$. We will explore this point, and some other
structures associated with constructing the unstable phase, in
\S\ref{z-tit-mid}.

It is interesting to see how the basic physical features of \ads{3} are actually encoded in s-L theory. For
instance, we would like to see how the Atick-Witten tachyon appears in the s-L background at exactly the correct
temperature. We describe the model perturbatively in $\lambda_2$ around a $sine-Liouville\times S^1$
($\lambda_1$ held fixed)
\begin{align}
    \Delta\mathcal{L}=\lambda_2\mathcal{V} =\lambda_2 V_{j,m,\bar m}\cos R_\varphi(\varphi_L-\varphi_R),
\end{align}
where $V_{j,m,\bar m}$ are the vertex operator of the 2-dim s-L theory (the $x-\phi$ plane) defined in
\S\ref{y-tit-comp-fzz}. The asymptotic behavior (large $\phi$) of the operator is,
\begin{equation}
    V_{j,m,\bar m}\xrightarrow{\phi\rightarrow+\infty}\frac{e^{i\sqrt{\frac{2}{k}}(m\chi_L-\bar m\bar\chi_R)}}{1+2j}
    \left[e^{2Qj\phi}+R_{j,m,\bar m}e^{-2Q(j+1)\phi}+\cdots\right].
\end{equation}
The coefficient $R_{j,m,\bar m}$ is a reflection coefficient of an incoming wave $e^{2Q j\phi}$ scattering off
the s-L interaction and returning as an outgoing wave $e^{-2Q(j+1)\phi}$. We are interested in a perturbation
which carries no winding and no momentum in the $x$ circle, hence we set $m=\bar m=0$ and it is enough for us to
note that $R_{j,0,0}\neq0$.\footnote{For a detailed expression for $R_{j,m,\bar m}$ and a comprehensive
explanation of the reflection phenomena in s-L see \cite{Aharony:2004xn}.}

Since we are interested in instabilities in the interior of the space, keeping its asymptotic form fixed, then
we are allowed to insert only vertex operators $\mathcal{V}$ that are normalizable (or delta-function
normalizable). In the asymptotic weak coupling region we need to check:
\begin{equation}
    \frac1{g_s^2}\mathcal{V}\xrightarrow{\phi\rightarrow+\infty}{\rm finite}.
\end{equation}
Using the asymptotic behavior and the linear dilaton slope we find
two conditions for normalizability
\begin{align}
    2 Q j + Q \leq 0
    &&
    -2 Q(j+1) + Q \leq 0.
\end{align}
The only solution to both constraints is ~$j=-\frac12+i s$ for some
real value of $s$. This is the familiar condition that the states be
delta-function normalizable in the linear dilaton asymptotics. From
the s-L point of view it is required that the dressed $\mathcal{V}$
is marginal or relevant, while $V_{j,0,0}$ is delta-function
normalizable. This imposes the following restriction on $s$:
\begin{equation}\label{sstts}
    \frac14R_\varphi^2-\frac{j(j+1)}{k-2} \leq 1\rar1
    R_\varphi\leq\sqrt{4-\frac{1+4s^2}{k-2}}.
\end{equation}
The smallest value of $\varphi$ radius where such and interaction term is possible is at $s=0$. Taking this
value and translating back to the coordinate conventions used for \ads{3} we get
\begin{equation}
    \beta^2 \leq \frac{4\pi^2}{k}\left(4-\frac{1}{k-2}\right)=\beta_H^2.
\end{equation}
This computation is quite suggestive. One can see many of the well
known features of \ads{3} space in the s-L theory. For example, the
fact one is forced to consider linear combinations of modes is much
easier to interpret in the s-L language (just because there is wall
from which there is reflection). In \ads{3}, the need to consider
linear combinations appears due to singularities of certain wave
functions in the interior. In addition, the calculation itself is
somewhat easier in the s-L language. Since there is a manifest
weakly coupled region near the boundary, dimensions can be
calculated using the zero point energy of the free theory.\footnote{Similar calculation can also be done in \ads{3}, but this requires some variable change which can actually be interpreted as T
duality (see \cite{Giveon:1998ns}).}

\subsection{The regulated model}\label{y-tit-comp-reg}

\tads3 and $BTZ$ with pure NS-NS fields are pathological in two
ways. The first is that the winding mode tachyon is delocalized and
can reach the boundary. The other is that they differ by a
non-normalizable mode. These pathologies are closely related to the
singularities of this special point in the moduli space of the
$D1-D5$ system which were discussed in \cite{Seiberg:1999xz}. Both
the "long strings" and the delocalization of the tachyon correspond
to the ability of long strings in thermal \ads{3} to escape to
infinity.\footnote{The two are actually more closely related. Both
long strings in Lorentzian space and the Atick-Witten tachyon in
\tads{} are described by a string winding one circle, and
quantization on an $S^1\times R$ worldsheet is very similar.}

Going to the T-dual picture, we have argued that the volume of this
torus is related to the asymptotic integral of the $B$ field (which
distinguishes our two phases and determines which circle is wrapped
by the delocalized tachyon). In other words, this non normalizable
mode is closely tied to the asymptotically flat potential for long
strings. If the system is regulated in a way that all the winding
states are confined to the bulk of space, than there is no
measurable difference near the boundary, even not by using extended
probes such as long strings. We conclude that if there exist
deformations which in effect trap such long strings, \tads3 and
$BTZ$ will differ by normalizable modes and would manifestly be in
the same ensemble under the usual Euclidean AdS/CFT rules.

Indeed, one can construct such examples. The $D1-D5$ CFT is singular
on a subspace of its moduli space \cite{Seiberg:1999xz} and one can
go off this subspace with a small deformation. We will rely on  the
B-field deformation constructed in \cite{Maldacena:1999mh,
Dhar:1999ax}. In the $D1-D5$ system, long strings which make their
way to the boundary correspond to instantonic strings on the
$\mathbb{T}^4$, which is wrapped by the $D5$ branes, shrinking to
zero size. This is a $D1$ brane which becomes point like and can
then leave the stack. Hence, one should prevent instantons from
shrinking. One way to do it is by turning on $B$ field on the
$\mathbb{T}^4$ which in effect modifies the moduli space of
instantons to that of a non-commutative $\mathbb{T}^4$. In the
non-commutative case, instantons cannot shrink due to the existence
of a new physical scale.

Taking the near-horizon of the regulated $D1-D5$ system
\cite{Maldacena:1999mh,Dhar:1999ax}  S-dualizing and Wick rotating
one obtains the background
\begin{align}\label{RR-regulated}
    &{ds'}^2_{str} = k
    \left[d\rho^2+\cosh^2\rho dt^2+\sinh^2\rho d\theta^2+d\Omega_3^2\right]
    +\sqrt{V}\left[\frac{(1-\epsilon)^2}{\cos^2\gamma}\,dT_2'^2+dT_2^2\right]\\
    &e^{-2\Phi'} = g_s^2\csp1
    C_{(2)}' = -2\frac{V}{g_s^2}\tan\gamma\,\varepsilon_{T_2'}\cr
    &F_{(5)}=2k\,\frac{\sqrt{V}\tan\gamma}{g_s^2}\frac{(1-\epsilon)^2}{\cos\gamma}(1+\star_{10})\left[
    \left(-id(\sinh^2\rho)\wedge dt\wedge d\theta+\frac1{\alpha'}\varepsilon_{\Omega_3}\right)\wedge\varepsilon_{T_2'}\right]\cr
    &dB_{(2)}'=2k (1-\epsilon)\left(-id(\sinh^2\rho)\wedge dt\wedge d\theta+\frac{1}{\alpha'}\varepsilon_{\Omega_3}\right)\cr
    &dT_2'^2 = dy_1^2+dy_2^2\csp1
    dT_2^2 = dy_1^3+dy_2^4\csp1
    \varepsilon_{T_2'}=dy_1\wedge dy_2
    \cr
    & 1-\epsilon=\left(1+\frac{V}{g_s^2}\tan^2\gamma\right)^{-1/2}\csp1 1>\epsilon\geq0.\cr
\end{align}
The limit $\gamma\rightarrow0$ (equivalently $\epsilon\rightarrow 0$) is the NS-NS Euclidean \ads3 solution. Consequently, the dynamics of a long string is now governed by this modified Nambu-Goto action . We can expand the action at large $\rho$ and read out the potential for the long string we studied before
\begin{gather}
    V_{F1} \sim \int d^2\xi\biggl(
   \frac12e^{2\rho}\left(1-\alpha\right)+const+O(e^{-2\rho})
    \biggr).
\end{gather}

Next we compactly the angles to find a deformation of \tads3 (or
$EBTZ$),
\begin{align*}
    t\cong t+\beta &&
    \theta\cong \theta+2\pi.
\end{align*}
It is already clear that the long strings no longer have a flat
potential. Rather, they are confined to the bulk of space. This was
the expected effect of this deformation. Hence, small deformations
of the singular point resolve the pathological non normalizable mode
which is measurable at infinity. Next carry out the same set of
T-dualities.\footnote{T-dualities with RR fields are discussed in
\cite{Fukuma:1999jt,Hassan:1999bv}. It is fortunate that in our case
the RR fields has no effect in the \ads{} directions enabling us to
use the familiar NS-NS duality formulas.} Writing only the NS-NS
fields and focusing on the \ads3 directions only we find the
regulated cigar,
%\begin{align}\label{RR-t-dual}
%    & ds^2  = k v(\rho) \left(d\rho^2+\tanh^2\rho d\chi^2 +d\varphi^2\right) \cr
%    & B_{(2)}' = 2ik v(\rho)\epsilon \sinh^2\rho \left[1+(1-\epsilon)\tanh^2\rho\right] d\varphi\wedge d\chi \cr
%    & g_s' = g_s\bigm/ v^2(\rho) \cr
%    &\varphi\cong \varphi+\beta ~,\qquad
%    \theta\cong \theta+2\pi \cr
%    & v(\rho) = \left[1-2\epsilon\sinh^2\rho+\epsilon^2\,\frac{\sinh^4\rho}{\cosh^2\rho}\right]^{-1}.
%\end{align}
\begin{align}\label{RR-t-dual}
    & ds^2  = k \left[d\rho^2+v(\rho)\left(\tanh^2\rho d\chi^2 +d\varphi^2\right)\right] \cr
    & B_{(2)}' = 2ik v(\rho)\epsilon\,\sinh^2\rho \left[1+(1-\epsilon)\tanh^2\rho\right] d\varphi\wedge d\chi \cr
    & g_s' = g_s\,\frac{v(\rho)}{\cosh^2\rho} \cr
    &\varphi\cong \varphi+\beta ~,\qquad
    \theta\cong \theta+2\pi \cr
    & v(\rho) = \left[1-2\epsilon\sinh^2\rho+\epsilon^2\,\frac{\sinh^4\rho}{\cosh^2\rho}\right]^{-1}.
\end{align}

We will focus on the case of $\beta$ close to $2\pi$. The technical
reason is the background \eqref{RR-t-dual} has a singularity at
finite $\rho$ which is purely an artifact of the T-duality. Around
$\beta\sim 2\pi$ we can truncate to leading order in $\epsilon$. We
see that the volume of the torus is rescaled by $\sim (1+2\epsilon
\tanh^2\rho)$. One can therefore glue, for the same background at
large values of $\rho$, an $\SL{2}/U(1)\times U(1)$ -like throat for
a range of values of this volume, which implies that \tads3 and
$EBTZ$ are in the same ensemble in this case.\footnote{The dilaton
needs to be adjusted at the point of the gluing which is possible.
Also, the value of $\int_{\mathbb{T}^2} B$ is an irrelevant
perturbation from the point of view of evolution in $\rho$ in the
cigar.}

\section{The middle point CFT}\label{z-tit-mid}

As discussed above, we expect (for intermediate temperatures) to
find a new fixed point, where both circles are on equal footing. In
this section we suggest what this theory might be, although more
work is needed to verify this picture. We will mainly work in the
Hawking-Page temperature \eqref{HP-temp}, where the two circles are
identical and the additional $\mathbb{Z}_2$ symmetry simplifies the
discussion, but will comment on the more general case.

We concentrate on  the unregulated model since we wish to use
worldsheet CFT arguments. When we go to the regulated model the
following changes occur
\begin{itemize}
\item Lines of fixed points in the unregulated model can collapse to a set
of discrete points as the spacetime regulator introduces a small
flow on this line. The origin of the small beta function is from the
region where the unregulated model is glued to a new space which
takes over as one approaches the boundary of spacetime.
\item RG flow which is infinite distance in coupling space turns into a
finite distance flow. The infinite distance flow is roughly similar
to the flow in \cite{Adams:2001sv} from $C/Z_N$ to $C$. Introducing
the regulator is the same as focusing on the flow from $C/Z_N$ to
$C$ in a finite region around the origin - such a region of
spacetime relaxes in a finite RG distance.
\end{itemize}
With these changes in mind we can discuss both cases at the same
time. The problem then boils down to the question of how one can
cap, at strong coupling (or in the IR in the AdS/CFT terminology), a
$linear-dilaton\times \mathbb{T}^2$ background. There are several
options, which can be located at different position along the flow.
In the following, we describe and evaluate these possibilities.

\medskip

{\bf 1. The maximally symmetric points}

The maximally symmetric plane of fixed points is characterized by
the fact that it has an $\SL{2,\Zf}$ discrete gauge symmetry which
acts on the parameters of the $\mathbb{T}^2$ (and no other
parameters transform under it), i.e., other than the shape of the
torus the symmetry between the different directions is preserved. As
usual it is natural to take the shape modulus of $\mathbb{T}^2$ for
this class of theories to live in the fundamental domain of
$SL(2,\Zf)$ (and of course there is still the volume modulus). The
only known candidate for this theory - and one that should be
considered since it has the right boundary condition to be included
in the ensemble - is $Liouville\times \mathbb{T}^2$. The cap is done
completely within the linear dilaton theory, and the $\SL{2,\Zf}$
acts as a symmetry on a decoupled $\mathbb{T}^2$. Considering this
theory, however, is problematic since, for the relevant values of
the linear dilaton slope, the coefficient $\alpha$ in the Liouville
interaction $e^{\alpha\phi}$ is imaginary and the interaction is
part of the delta-function normalizable spectrum.

If such a theory can be defined, then it is defined for any value of
$\mu_L$ - the coefficient of the Liouville interaction on the
worldsheet. Usually, this value is considered inconsequential as it
can be changed by shifting the linear dilaton coordinate $\rho$.
However, in our case one has to keep it as a modulus of the theory
since we fix the UV cut-off during a computation (at large value of
$\rho$).\footnote{ Phrased in another way, we fix the value of $g_s$
at the cut-off point.} $\mu_L$ then measures the length of the
throat from this cut-off. This gives us a line of fixed points. When
fixing the boundary conditions and summing over all bulk geometries
we have to sum over this line of fixed points.

It is important to emphasize that even if the CFT on the sphere is
well defined for all values of $\mu_L$, the range around $\mu_L=0$
is problematic because then the coupling is strong near the cap and
one needs to evaluate higher loops and non-perturbative effects. In
particular, if the integral over $\mu_L$ diverges for $\mu_L\sim 0$,
these strong coupling effects are dominant. The point $\mu_L=0$
itself - i.e. $linear-dilaton\times \mathbb{T}^2$ - certainly does
not exist as a perturbative String background. One can estimate
whether the behavior near $\mu_L\sim 0$ is indeed problematic, since
then the throat is long (even in the regulated model) and the
$\mu_L$ dependence of correlation functions dominated by the throat
is governed by a KPZ scaling similar to \eqref{kpzscl}. We leave
this to future work.

The suggestion here is in the same spirit as in
\cite{Giveon:2006pr}, which treats the Horowitz-Polchinski
correspondence principle \cite{Horowitz:1997jc,Horowitz:1996nw} from
a worldsheet point of view (for a specific class of black holes). In
that case, the near horizon of a black hole, in the vicinity of the
black hole/excited string phase transition, is described by an
$\SL{2}/U(1)$ cigar, and all angular information about the black
hole disappears. Here we also obtain a similar throat but keep an
additional $S^1$ which can encode angular information (since the
chemical potential is encoded in the shape of the $\mathbb{T}^2$).

\medskip

{\bf 2. A partially symmetric point}

In the language of the Euclidean instantons described in \S\ref{y-tit-rev-eucl} the transformation that interchanges $EBTZ$ and \tads3 , is $\tau' = -\frac1{\tau}$. At the Hawking page temperature $\tau= i$ and the transformation is a $\Zf_2$ symmetry of the boundary. The symmetry is broken in the interior of space by the choice of which circle contracts. One can ask whether there is a symmetric point which respects this $\Zf_2$ symmetry.

Our conjecture for the $\Zf_2$ symmetric point is to take \eqref{two-defor} and set $\lambda_1=\lambda_2$ (remember that $R_x=R_\phi$ in this point of moduli space), i.e.,
\begin{equation}\label{double-SL}
    \lambda e^{b\phi}\left(\cos R(x_L-x_R)+\cos R(\varphi_L-\varphi_R)\right),
\end{equation}
where $b$ is determined by fixing the scaling dimension and $R$ is
the radius obtained using FZZ duality. $SL(2,\Zf)$ is still a
discrete gauge symmetry but now it acts on the shape modulus of the
$\mathbb{T}^2$ and on the two cycles on which we chose to turn on
the s-L interaction. The fundamental domain of this theory is now
larger than the fundamental domain of $SL(2,\Zf)$.

We would like to emphasize again that this is a conjecture, as it is
not clear that this theory exists. For examples the techniques of
\cite{Kazakov:2000pm} cannot be applied in any simple way. Hence, we
will have to argue its existence indirectly as follows.

\begin{figure}[ht]
  \begin{center}
      \subfigure[\label{F1}]{\scalebox{0.45}{\includegraphics[width=1\textwidth]{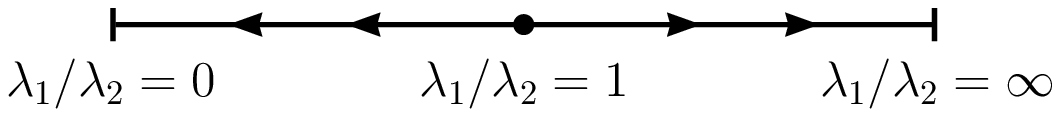}}} \\
      \mbox{
      \subfigure[\label{F2}]{\scalebox{0.45}{\includegraphics[width=1\textwidth]{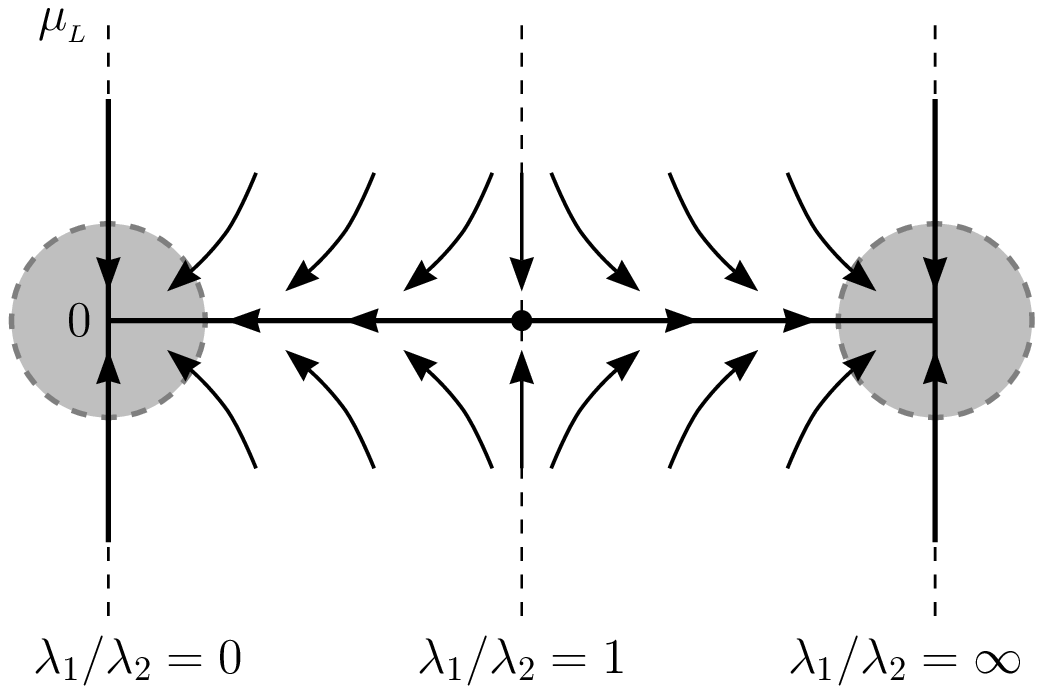}}} \quad
      \subfigure[\label{F3}]{\scalebox{0.45}{\includegraphics[width=1\textwidth]{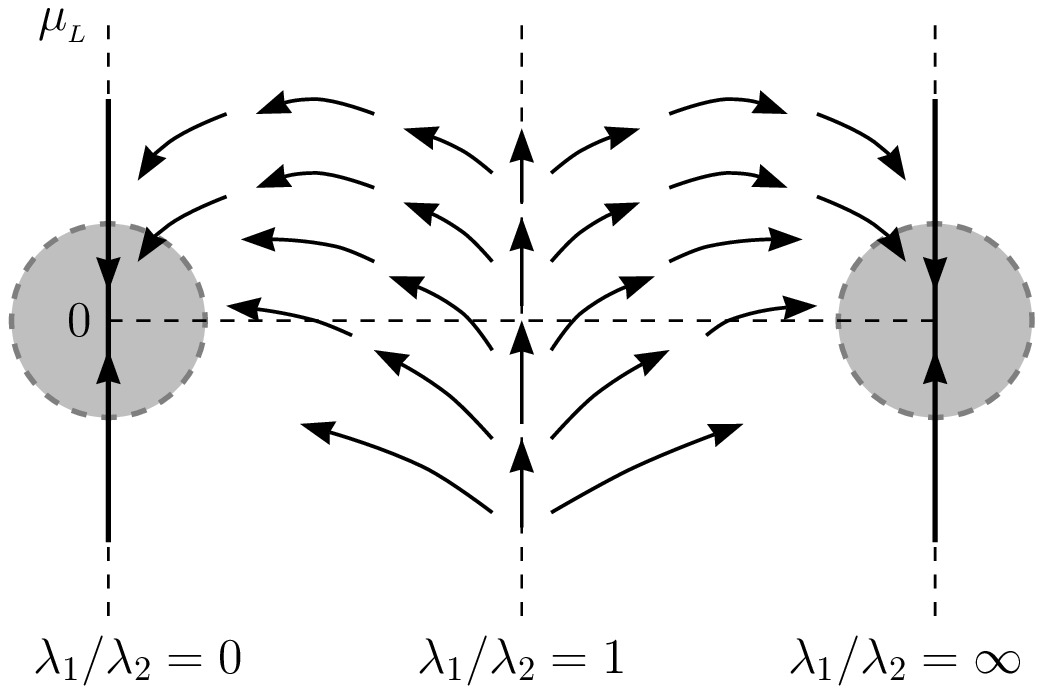}}} \quad
      }
      \caption{The three possible flow patterns at the HP temperature. Figure (a) is the naive one dimensional flow in $\frac{\lambda_1}{\lambda_2}$. Figure (b) is the conjectured flow case when we include a small $\mu_L$. Figure (c) is an example of what could go wrong. In the latter $\lambda_1=\lambda_2$ is not a fixed point }
  \end{center}
\end{figure}

If we can restrict our attention to the two operators in
(\ref{two-defor}), which we can for $\lambda_1\ll\lambda_2$ (or
$\lambda_2\ll\lambda_1$), then the existence of an unstable fixed
point at $\lambda_1=\lambda_2$ is guaranteed. The flow is then
depicted in figure \ref{F1}. However, it is not clear that one can
restrict to only these two operators. If we follow
\cite{Kazakov:2000pm} then there are at least three operators which
play a role. In addition to the two sine-Liouville interactions in
\eqref{two-defor}, one also expects that a Liouville interaction
could be turned on (we will denote the coefficient of the latter
$\mu_L$ here as well). If $\lambda_1=0$ (or $\lambda_2=0$) FZZ
duality conjectures that we can set $\mu_L=0$ (as was shown
explicitly for $k=9/4$ in \cite{Kazakov:2000pm}). For $\mu_L$ very
small, such that the Liouville cap is behind the sine-Liouville i.e.
deeper into the strong coupling area, we expect $\mu_L$ to be
irrelevant. This describes the shaded areas in figure \ref{F2}. Note
however that close to the entire $\mu_L=0$ line (i.e.
$\mu_L<<max(\lambda_1,\lambda_2)$) the Liouville interaction cap is
behind the sine-Liouville combined cap. Hence we expect that the
Liouville interaction will always be irrelevant for small enough
$\mu_L$, which gives the entire \ref{F2} phase diagram and an
unstable fixed point at $\lambda_1=\lambda_2,\ \mu_{Liouville}=0$.

These arguments implicitly assume that certain orders of limit do not matter (for example if we follow \cite{Kazakov:2000pm}, which is applicable only for $k=9/4$ to start with, where one starts with small $\lambda$ and resums the expansion in this parameter) and that no additional operators can appear in front of the s-L caps. Hence it is suggestive, but not a rigorous argument.

\medskip

It is also interesting to see what other middle point CFTs can be
obtained by using other caps - i.e. by taking, in the asymptotic
$linear\ dilaton\times \mathbb{T}^2$ some other operators from the
$\mathbb{T}^2$ and dressing them by an operator from the linear
dilaton, which grows in the strong coupling region. The most natural
example would be to take the Poincar\'e patch.\footnote{This analysis
was also carried out by S. Minwalla.} We begin with the following
Euclidean metric, which is a solution of GR,
\begin{align}
    &ds^2=k\left(d\rho^2+\frac{e^{2\rho}}{4}d\theta^2+\frac{e^{2\rho}}{4}dt^2\right)
    \cr
    &B_{(2)}=-\frac{ik}{2}\left(e^{2\rho}-2\right)dt\wedge d\theta
    \cr
    &e^{\Phi}=g_s
    \cr
    &t\cong t+\beta~,\quad \theta\cong\theta+2\pi.
\end{align}
In the above we kept the constant term in the B-field, which plays an important role in our backgrounds, as is clear by now. This solution has the same symmetry pattern of the Liouville theory. We have chosen this specific B field since it is the one that corresponds to the Poincar\'e patch. However, as far as the behavior at infinity is concerned we could change this value (while staying in the ensemble or in the regulated model).

Next we would like to go over the $linear\ dilaton\times
\mathbb{T}^2$ picture. Carrying out the sequence of T-dualities as
before we find
%T-dualizing the background in the $t$ direction we find,
%\begin{align}
%\begin{split}
%    &ds^2=k\biggl[d\rho^2
%    +\left(1-e^{-2\rho}\right)\left(d\theta-id\tilde t\right)^2
%    +2i e^{-2\rho}d\tilde t\left(d\theta-id\tilde t\right)+\cr
%    &\qquad\qquad
%    +\left(1+e^{-2\rho}\right)d\tilde t^2
%    \biggr]
%\end{split}
%    \cr
%    &B_{(2)}=0
%    \cr
%    &\Phi=\log g_s-\log\left(\frac{k\beta^2}{16\pi^2}\right)-2\rho
%    \cr
%    &\tilde t\cong \tilde t+\frac{4\pi^2}{\beta k}~,\quad \chi\cong\chi+2\pi.
%\end{align}
%We define the new coordinate $\chi = \theta-i\tilde t$~ and T-dualize in the $\tilde t$ direction to find the background,
\begin{align}\label{middle-back}
    &ds^2=k\left(d\rho^2+\frac{d\chi^2+d\varphi^2}{1+e^{-2\rho}}\right)
    \cr
    &B_{(2)}=\frac{2ik}{\left(1+e^{2\rho}\right)}d\varphi\wedge d\chi
    \cr
    &\Phi=\log g_s
    +\log\frac4{1+e^{2\rho}}
    \cr
    &\varphi\cong \varphi+\beta~,\quad \chi\cong\chi+2\pi.
\end{align}

The geometry far away at the weak coupling $\rho\rightarrow\infty$
region indicates a cap made out of the volume and B-field of the
$\mathbb{T}^2$, dressed by a profile in the Liouville direction. At
$\rho\rightarrow -\infty$ the two circles shrink while the B-field
and coupling go to a constant value.

Since the circles shrink exponentially, it is not clear how to
analyze this theory. We would like however to point to the
possibility that this theory flows to a Liouville theory. The
shrinking $\mathbb{T}^2$ could either disappear from the theory (as
is the case in \cite{Giveon:2006pr} for an $S^2$) or it can
stabilize at some finite stringy radii (since here, unlike $S^2$,
there is a CFT for any radii of $\mathbb{T}^2$) but in any case,
from considering the central charge, a linear dilaton term should be
generated. The perturbation at infinity then has the right quantum
numbers to mix into a Liouville wall. We can perhaps separate the
Liouville wall from where the $\mathbb{T}^2$ shrinks by changing the
value of the $B$-field.

%
%is of $\mathbb{T}^2$ shrinking as $\rho\rightarrow-\infty$. This is
%depicted schematically in figure \ref{xtheoryfig}. The solution has
%the expected $\mathbb{Z}_2$ symmetry if $\beta=2\pi$. It is tempting
%to associate this solution with our double s-L theory via some
%generalized FZZ duality for $c=2$ matter. Note that near the weakly
%coupled region, (\ref{middle-back}) is exactly a sum of two cigar
%like deformations of the action, with exactly the same coefficients.
%This is again a reminiscent of the procedure we employed in the "FZZ
%dual" case. Needless to say the central charge of this solution is
%exactly the same as of the stable points (since the dilaton slope at
%the weakly coupled region is the same). Note that on top of the two
%cigar like deformations at infinity, there are more small
%corrections in the bulk. In particular, there is a small $H$ flux
%needed for conformal invariance. We conclude that in both
%presentations, we have a natural candidate for the unstable fixed
%point with $\Zf_2$ symmetry.

%\begin{center}
%\EPSFIGURE{xtheory.eps, width=0.35\textwidth}{\label{xtheoryfig} The topological structure of our solution which is $\mathbb{Z}_2$ symmetric.}
%\end{center}
%\begin{figure}[ht]
%  \begin{center}
%      \includegraphics[width=0.35\textwidth]{xtheory.eps}
%    \caption{The topological structure of our solution which is $\mathbb{Z}_2$ symmetric.}
%    \label{xtheoryfig}
%  \end{center}
%\end{figure}
%\medskip

We have presented our solutions only at the HP temperatures, but
similar caps can occur for other values of the temperature as well.
This allows testing our proposal by approaching the Hagedorn
temperature of \ads3 from below. As in figure \ref{pot1}, in this
regime the two fixed points should be close to each other, and one
can hope to perform a perturbative calculation detecting the nearby
fixed point close to \ads3. This would be especially interesting in
the s-L language, where one can test these ideas explicitly.

We should mention that the role of the unstable fixed point in the case of \ads{5} is played by the small black hole which can decay to a larger black hole (which is stable at high temperatures) or a gas of particles in \ads{5} (which is stable at low temperatures). In some sense, our solution is a small black hole as well since the time circle is contractible and the horizon has exactly zero area.

\subsection{Summary of the flows}

Let us briefly summarize the flows.\footnote{A boundary CFT
interpretation of the phases in \ads{3} was proposed in
\cite{Kurita:2004yn}.} Part of this picture can be derived from
considering the free energy in the dual field theory, but now one
can be more precise about the different CFT's in the different
regimes.

\begin{figure}[ht]
  \begin{center}
    \includegraphics[width=.8\textwidth]{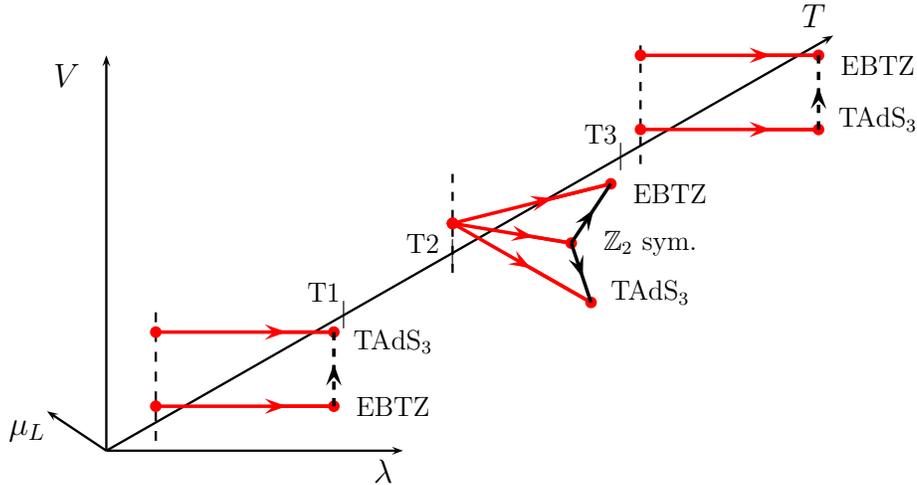}
    \caption{Qualitative picture of the flow between the various
    theories}
    \label{fig:QFT-flow}
  \end{center}
\end{figure}

The qualitative flow between the CFT's is described in figure
\ref{fig:QFT-flow}.
%\begin{center}
%\EPSFIGURE{QFT_flow.eps,width=.7\textwidth}{\label{fig:QFT-flow}RG flow diagram for the 2-dim QFT describing thermal \ads3. The blue line is a line of unstable fixed points.}
%\end{center}
%\begin{center}
%\EPSFIGURE{GrandFlow.eps,width=.7\textwidth}
%{\label{fig:QFT-flow}RG flow diagram as a function of temperature in the 2-dim parameter space $V$ and  $\lambda_{s-L}$, the volume of the torus and the sine-Liouville coupling. The fourth axis is the Liouville interaction $\mu$, the plot is constrained to the $\mu=0$ plain. The temperature axis is divided to quadrants by the critical temperatures (HP, Hagedorn) of in figure \ref{fig:phase-diagram}.  The red (solid) flow lines are regular flows in the s-L plain while the black (dashed) flow lines are flows related to tachyon condensation generating a change of the boundary conditions. In the unique case of the HP temperature both type of flows are regular with no change of the boundary conditions.}
%\end{center}
In this diagram the axis that goes into the page is the temperature
axis. The four temperatures regimes, separated by temperatures $T1$
,$T2$ and $T3$ refer to those in figure\ref{fig:phase-diagram}. This
diagram applies for the unregulated model, and we discussed before
how to pass from it to the flow picture for the regulated model.

The relevant axis in the diagram, in addition to the temperature
axis, are $\lambda$ which stands for both sine-Liouville
interactions, $V$ which denotes the volume of the $\mathbb{T}^2$ at
infinity and $\mu_L$ which is the coefficient of the Liouville
interaction. We will be interested in the regime where the Liouville
wall is behind the s-L walls, hence we can set $\mu_L=0$ in the
discussion. There are 3 types of arrowed lines - red arrowed lines
denote using a marginal s-L interaction to deform the theory (and
FZZ duality to write it geometrically). This is not really a flow
but a finite distance change as far as the worldsheet CFT is
concerned. Solid black arrowed lines denote ordinary CFT flow.
Dashed black arrowed lines denote an infinite distance flow which
can also change the boundary conditions on the volume of the
$\mathbb{T}^2$. Vertical unarrowed dashed lines simply indicate
lines of different V, keeping all other parameters fixed.

The figure maps the different theories at 3 different temperatures:
The HP temperature, at a high temperature (above the Hagedorn
temperature) and at a low temperature (below the dual Hagedorn
temperature, where the $EBTZ$ is tachyonic). The latter two are of
course images of each other under the exchange of $EBTZ$ with \tads3
and $\beta$ with inverse $\beta$.

At the HP temperature we can turn on a s-L interaction on either
circles to go either $EBTZ$ or \tads3. We also expect that there is
a symmetric point where both caps are turned on. From it, one can
flow to either $EBTZ$ or \tads3.

At temperatures above the Hagedorn temperature we can go to either
$EBTZ$ or \tads3 using an s-L interaction, but one needs to start
with different values of $V$. This is depicted by the red line
starting at different values of $V$ at $\lambda=0$. The flow from
\tads3 to $EBTZ$ via the Atick-Witten tachyon has to change the
value of $V$.

Tachyons can change the boundary conditions of non-compact CFTs
under some circumstances. The most straightforward case are tachyons
which are completely delocalized. In this case one can condense
their zero mode (the most familiar case is that of the bosonic
theory) and the entire space changes, including the behavior at
infinity. Such a flow is expected to change the central charge by
the Zamolodchikov c-theorem \cite{Zamolodchikov:1986gt}. The other
case is if they are localized but mix into the delta function
normalizable states of the non-compact CFTs. In this case a shell
expands from where the tachyon is localized to the boundary but,
strictly speaking, it does not reach it in finite distance in
coupling space. This is what happens in \cite{Adams:2001sv} and in
such cases the central charge does not have to change.

The situation here is intermediate. As we saw before we can identify
the tachyon in the delta function normalizable spectrum in
$sine-Liouville\times S^1$. If we build a wave packet out of it then
we have a kind of localized tachyon, which can mix into the delta
function normalizable states of the volume mode (which are not
necessarily tachyonic). The latter will make up the shell that
propagates to infinity of spacetime and eventually changes, in
infinite distance in coupling space, the asymptotic volume of the
$\mathbb{T}^2$. This picture is supported by the regulated model
where the flow is a localized tachyon, finite distance, flow which
changes only a region of space (very similar to $AdS_5\times S^5$).
As the regulator is removed the length of the region which changes
grows to infinity until it covers the entire space suggesting an
infinite distance flow in which the original tachyon mixes with the
modes described above of the volume.

%Not building a wave packet but condensing the $s=0$ mode is more puzzling (see the discussion around \eqref{sstts}), and we will a full discussion to future work. Let us point out however that it is not clear that this is possible. In situation where there is a delta-function normalizable spectrum of tachyonic operators because there is translation invariance in a non-compact direction one can discuss the condensation of the zero mode because then this non-compact direction decouples and one discusses a discrete operator in the remaining CFT. This is not the case here since there is no translation invariant the tachyon profile is $\rho$ dependent. In fact, if one wants to take the back-reaction of this field (or any wave packet) then the back-reaction is suppressed at large values of $\rho$ due to the decrease in $g_s$ suggesting again localized wave packet construction.

Not building a wave packet but condensing the $s=0$ mode is more
puzzling (see the discussion around \eqref{sstts}), and we postpone
a full discussion to future work. Let us point out however that it
is not clear that discussing only this mode makes sense. In
situations where a delta-function normalizable spectrum of tachyonic
operators arise due to a translation invariance in a non-compact
direction one can safely discuss the condensation of the zero mode
because then the non-compact direction decouples and one discusses a
discrete operator in the remaining CFT. This is not the case here
because there is no translation symmetry in $\rho$ and the tachyon
profile is $\rho$ dependent. In fact, if one wants to take the
back-reaction of this field (or any wave packet) then the
back-reaction is suppressed at large values of $\rho$ due to the
decrease in $g_s$ suggesting again the localized wave packet
construction.

\section{Discussion and open questions}\label{z-tit-open}

We suggested a unified description for geometric and tachyonic
capping in String theory, and applied it to the study of the
\tads3/$EBTZ$ phase diagram where it turns out to be a useful tool
for understanding the intermediate worldsheet CFTs. We expect that
this technique of evaluating competing tachyon condensation, would
be useful in other topology change problems in String theory and GR.
Here we used the symmetries of the problem, such as the $\Zf_2$
symmetry which interchanges the two circles, to argue the existence
of a new fixed point. But more generally one expects similar fixed
points at the end of the separatrix that separates between the phase
in which one tachyon dominates to the phase in which another does.
This entails the comparison of different "walls" made out of
Liouville-like interactions times different operators from some
internal compact CFTs.

A concrete open problem is to provide more evidence for the
existence of the new conformal field theory which we conjectured to
describe the unstable phase. There are many tests one can make to
check this conjecture, some of which were described in
\S\ref{z-tit-mid}. Another open problem is that the hard dynamical
part of the process - i.e., the details of the RG flows - remains to
be understood. Particularly interesting are the flows away from the
HP temperature due to the need to change a non-normalizable mode.
This problem occurs in the theory because the spacetime CFT is
singular, and hence the UV/IR relation in AdS$_{3,{\rm NS-NS}}$ is
rather unusual. In the regulated model, in which RR charges are
turned on, the UV/IR relation is the standard one, but this model is
not solvable.  We plan to pursue these checks in the future.

This project started when attempting to study the Lorentzian $BTZ$
as a global model which realizes the time dependent Misner space or
Grant space. %, and study the condensation of strings in the whiskers of these models.
However, already the Euclidean $BTZ$/\tads3
exhibits a set of poles in the partition functions which are
practically identical to the poles encountered in Misner/Grant
spaces \cite{Cornalba:2002fi,Pioline:2003bs}. In \ads3 these poles disappear when
turning on a RR fields - it would be very interesting to find out
what is the corresponding deformed model in the Misner/Grant cases.
%More generally, the Lorentzian black hole remains poorly understood
%at the level of exact CFT. It would be very interesting to
%understand the latter and map the phase diagram of the spacetime CFT
%in the microcanonical ensemble.

\section{Acknowledgments}\label{z-tit-ack}

We would like to thank A.~Giveon, E.~Rabinovici, M.~Rangamani S.~Razamat,
M.~Rozali, Y.~Sekino, S.~Shenker, V. Shpitalnik, A.~Yarom and in
particular Ofer Aharony, Shiraz Minwalla and Eva Silverstein for
illuminating and useful discussions. Z.K would like to thank
Stanford Institute for Theoretical Physics and SLAC for their
hospitality during final stages of this project.

This work is supported by the Israel Science Foundation Center of
Excellent program (grant number 1468/06), by the EU RTN networks
program, by the German-Israeli Foundation for Scientific Research
and Development, by Minerva, by the Einstein Center, by the
Blumenstein foundation and by a grant of DIP (H.52).

\appendix

\section{Mini-Superspace analysis of the winding string}\label{z-tit-min}

The Atick-Witten tachyon discussed in \S\ref{z-tit-hag} is a winding mode around the Euclidean time circle
\cite{Atick:1988si}. We apply a minisuperspace quantization of the string in thermal \ads3 to study the
properties of this mode. We use bosonic String theory and ignore the existence of the usual bulk tachyon in the
spectrum. The action of a string in thermal \ads3 is,
\begin{gather}
    S=\frac{k}{4\pi}\int d^2\sigma\bigl[(\partial_a\rho)^2+\cosh^2\rho(\partial_at)^2+\sinh^2\rho(\partial_a\theta)^2
    -2\sinh^2\rho\epsilon_{ab}\partial_at\partial_b\theta\bigr],
\end{gather}
where $a$ and $b$ are worldsheet vector indices, $\epsilon_{ab}$ is a 2-dim antisymmetric tensor. The target
space coordinates match those of \eqref{TADS-background}. We consider a minisuperspace ansatz \footnote{It is
easy to check that the ansatz is competent with the equation of motion.} for the winding mode,
\begin{align}
    t= t(\s^2)+\frac{w\beta}{2\pi}\s^1
    &&
    \rho = \rho(\s^2)
    &&
    \theta = \theta(\s^2)
    \gsp2, w\in\mathbb{Z}.
\end{align}
The minisuperspace Lagrangian is,
\begin{gather}\label{minis-action}
    S=\frac{k}{2}\int
    d\sigma^2\biggl[\dot{\rho}^2+\cosh^2(\rho)\dot{t}^2+\sinh^2(\rho)\biggl(\dot{\theta}-\frac{w\beta}{2\pi}\biggr)^2
    +\biggl(\frac{w\beta}{2\pi}\biggr)^2\biggr],
\end{gather}
the dot stands for derivatives in respect of the worldsheet coordinate $\s^2$. A canonical quantization of
the Euclidean action follows \cite{Polchinski:1998rq} (chapter 8),
\begin{align*}
    v^n=i\dot{X^n}&&
    P_n=-\frac{\partial\mathcal{L}}{\partial v^n}&&
    H=\mathcal{L}+P_nv^n.
\end{align*}
Applied to the minisuperspace action \eqref{minis-action} for winding $w=1$ we find the Hamiltonian \footnote{It
is important to take care of ordering ambiguities in the Hamiltonian. In the case at hand, all ambiguities are
fixed by the existence of a unique quadratic differential consistent with the symmetries.}
\begin{multline}\label{mini-fullHam}
    H=\frac{1}{2k}\left(-\partial_\rho^2-\frac{\partial_\rho\left(\sqrt{g}\right)}{\sqrt{g}}\partial_\rho\right)
    +\frac{(P_t)^2}{2k\cosh^2\rho}+
    \\+
    \frac{1}{2k\sinh^2\rho}\left(P_{\theta}+\frac{ik\beta}{2\pi}\sinh^2\rho\right)^2
    +\frac{k}{2}\cosh^2\rho\left(\frac{\beta}{2\pi}\right)^2+\frac{a}{2k},
\end{multline}
with $\sqrt{g}=\frac12\sinh(2\rho)$. The additional constant $a$ comes by computing correctly the zero point
energy of all modes that have been integrated out (this will be carried out in the sequel). The imaginary term
in the Hamiltonian is a consequence of the imaginary $B$-field. For the zero momentum case $P_\theta=P_t=0$ the
imaginary part vanishes and the eigenvalue equation simplifies. The regularity of the wave function on the
disc $\rho,\theta$ imposes a simple constraint at $\rho=0$
\begin{gather}
    \pdf{\Psi(\rho)}{\rho}\bigm|_{\rho=0} = 0,
\end{gather}
and the eigenvalue problem is
\begin{equation}\label{mini-eigenvalue}
    E\Psi(\rho)=\frac{1}{2k}\left[-\partial_\rho^2-2\coth(2\rho)\partial_\rho
    +\left(\frac{\beta k}{2\pi}\right)^2+a\right]\Psi(\rho).
\end{equation}
It is interesting that the potential energy for this winding mode is exactly constant (due to cancelation of the
winding energy by the $B$-field coupling). The solutions of this eigenvalue problem are well known. Their
asymptotic form and the exact eigenvalue set are \footnote{We ignore the shift in $k$ which is invisible in the
simple reduction of the model we employ here.}
\begin{align}\label{mini-wf}
    \Psi_j(\rho) \sim \exp\left(2j\,\rho\right)
    &&
    E_j = \frac1{2k}\left[-4j(j+1)+\left(\frac{\beta k}{2\pi}\right)^2+a\right].
\end{align}
Remembering that there is a measure factor $\sqrt{g}\sim e^{2\rho}$ in the norm formula, we arrive at the
conclusion that the allowed set of $j$'s is $j=-\frac12+is$ where $s$ is any real number. For other cases the
wave function is either not normalizable at infinity or singular at the origin. Thus only continuum normalizable
solutions exist (which in particular means that these winding modes are not strictly localized). To reproduce
the Hagedorn temperature as well as the relevant terms in the one loop partition function, we need to compute
$a$, which we do in the following subsection.

\subsection{Calculation of the zero point constant}\label{y-tit-min-a}
To perform this computation one should examine more carefully the structure of the full CFT, using current
algebra techniques. Following \cite{Kutasov:1999xu}, the worldsheet stress tensor of $\mathrm{H}_3^+$ is
expressed in terms of affine $\SL{2,\Rf}\times\SL{2,\Rf}$ algebra currents,
\begin{align}
    T^{ws}(z)=\frac1{k-2}\left(-(J^3)^2+J^+J^-\right)
    &&
    \bar T^{ws}(\bar z)=\frac1{k-2}\left(-(\bar J^3)^2+\bar J^+\bar J^-\right).
\end{align}
The currents obey the OPE,
\begin{align}
    &J^3(z)J^{\pm}(w)\sim \frac{\pm J^{\pm}(w)}{z-w}
    &&\bar J^3(\bar z)\bar J^{\pm}(\bar w)\sim \frac{\pm \bar J^{\pm}(\bar w)}{\bar z-\bar w}
    \cr
    &J^-(z)J^+(w)\sim \frac{k}{(z-w)^2}+\frac{2J^{3}(w)}{z-w}
    &&\bar J^-(\bar z)\bar J^+(\bar w)\sim \frac{k}{(\bar z-\bar w)^2}+\frac{2\bar J^{3}(\bar w)}{\bar z-\bar w}
    \cr
    &J^3(z)J^3(w)\sim \frac{-k/2}{(z-w)^2}
    &&\bar J^3(\bar z)\bar J^3(\bar w)\sim \frac{-k/2}{(\bar z-\bar w)^2}.
\end{align}
\tads3 with parameter $\tau$ is an orbifold of $\mathrm{H}_3^+$. The orbifold generators twist the currents
\begin{equation}
    \left(J^3,~J^+,~J^-\right)\longrightarrow
    \left(J^3,~e^{-2\pi i\tau}J^+,~e^{+2\pi i\tau}J^-\right).
\end{equation}
These monodromy conditions allow us to write the oscillator expansion of the currents in the $n$-th twisted
sector
\begin{align}
    J^3(z) = \sum_{m}\frac{J^3_m}{z^{m+1}}
    &&
    J^\pm(z) = \sum_{m}\frac{J^\pm_{m\mp n\tau}}{z^{m+1\mp n\tau}}.
\end{align}
Using standard CFT techniques, we can derive the twisted commutation relations
\begin{align}\label{twisted algebra gen}
    [J^3_m,J^{\pm}_{l\mp n\tau  }]=&\,\pm J^{\pm}_{m+l\mp n\tau}\cr
    [J^-_{m+ n\tau },J^+_{l- n\tau }]=&\,2J^3_{m+l}+\delta_{m+l} k(m+n\tau)\cr
    [J^3_m,J^3_l]=&\,-m\frac k2 \delta_{m+l}.
\end{align}

From here, one has all the information needed to compute the minisuperspace Hamiltonian with the correct zero point
energy. The ambiguity previously present is now resolved by calculating the normal ordering of all the higher
string modes and regularizing the infinite sum in a way consistent with the Virasoro algebra. We list the
contributions to the total zero point energy in the following :
\begin{itemize}
    \item ghosts contribute $2/12$
    \item the unitary CFT $\mathcal{M}$ adds $\frac1{12}\left(\frac{6}{k-2}-23\right)$
    \item from the Klein-Gordon equation (zero modes) we get $\frac{1}{2(k-2)}$
    \item ordering of higher modes (using zeta function regularization) $-\frac{k}{4(k-2)}$
    \item the length of the string gives $\frac12k(\beta/2\pi)^2$
\end{itemize}
Summing up all the contribution we find the zero point energy
\begin{equation}
    \frac12k(\beta/2\pi)^2-2+\frac1{2(k-2)}.
\end{equation}
This vanishes exactly at the Hagedorn temperature \eqref{hagedorn_temp}. One can do a little more matching
(the zero mode) pieces of the exact partition function \eqref{partition-strip} by calculating the partition
function of the minisuperspace model
\begin{gather}
    \sum e^{-2\pi\tau_2 H}=\sum_{m\in \mathbb{Z}}\int_{s\in\mathbb{R}} ds e^{-2\pi i
    \tau_2m\beta}e^{-2\pi\tau_2\frac{2s^2}{k-2}}e^{-2\pi\tau_2\left(\frac12k\beta^2-2+\frac1{2(k-2)}\right)}.
\end{gather}
The poles are reproduced from the sum over imaginary energies, as happens in some time dependent backgrounds
\cite{Pioline:2003bs} (see \cite{Berkooz:2002je} for introduction to these time dependent models).The power $\frac1{\sqrt{\tau_2}}$ in the partition function is a consequence of summing over a
continuum of states, as expected.


\begin{thebibliography}{99}


%\cite{Maldacena:1997re}
\bibitem{Maldacena:1997re}
  J.~M.~Maldacena,
  ``The large N limit of superconformal field theories and supergravity,''
  Adv.\ Theor.\ Math.\ Phys.\  {\bf 2}, 231 (1998)
  [Int.\ J.\ Theor.\ Phys.\  {\bf 38}, 1113 (1999)]
  [arXiv:hep-th/9711200].
  %%CITATION = IJTPB,38,1113;%%


%\cite{Witten:1998qj}
\bibitem{Witten:1998qj}
  E.~Witten,
  ``Anti-de Sitter space and holography,''
  Adv.\ Theor.\ Math.\ Phys.\  {\bf 2}, 253 (1998)
  [arXiv:hep-th/9802150].
  %%CITATION = 00203,2,253;%%

%\cite{Gubser:1998bc}
\bibitem{Gubser:1998bc}
  S.~S.~Gubser, I.~R.~Klebanov and A.~M.~Polyakov,
  ``Gauge theory correlators from non-critical string theory,''
  Phys.\ Lett.\  B {\bf 428}, 105 (1998)
  [arXiv:hep-th/9802109].
  %%CITATION = PHLTA,B428,105;%%


%\cite{Witten:1998zw}
\bibitem{Witten:1998zw}
  E.~Witten,
  ``Anti-de Sitter space, thermal phase transition, and confinement in  gauge
  theories,''
  Adv.\ Theor.\ Math.\ Phys.\  {\bf 2}, 505 (1998)
  [arXiv:hep-th/9803131].
  %%CITATION = 00203,2,505;%%


%\cite{Hawking:1982dh}
\bibitem{Hawking:1982dh}
  S.~W.~Hawking and D.~N.~Page,
  ``Thermodynamics Of Black Holes In Anti-De Sitter Space,''
  Commun.\ Math.\ Phys.\  {\bf 87} (1983) 577.
  %%CITATION = CMPHA,87,577;%%

%\cite{Barbon:2001di}
\bibitem{Barbon:2001di}
  J.~L.~F.~Barbon and E.~Rabinovici,
  ``Closed-string tachyons and the Hagedorn transition in AdS space,''
  JHEP {\bf 0203} (2002) 057
  [arXiv:hep-th/0112173].
  %%CITATION = JHEPA,0203,057;%%

%\cite{Barbon:2002nw}
\bibitem{Barbon:2002nw}
  J.~L.~F.~Barbon and E.~Rabinovici,
  ``Remarks on black hole instabilities and closed string tachyons,''
  Found.\ Phys.\  {\bf 33} (2003) 145
  [arXiv:hep-th/0211212].
  %%CITATION = FNDPA,33,145;%%
  
%\cite{Barbon:2004dd}
\bibitem{Barbon:2004dd}
  J.~L.~F.~Barbon and E.~Rabinovici,
  ``Touring the Hagedorn ridge,''
  arXiv:hep-th/0407236.
  %%CITATION = HEP-TH/0407236;%%

%\cite{Horowitz:2006mr}
\bibitem{Horowitz:2006mr}
  G.~T.~Horowitz and E.~Silverstein,
  ``The inside story: Quasilocal tachyons and black holes,''
  Phys.\ Rev.\  D {\bf 73}, 064016 (2006)
  [arXiv:hep-th/0601032].
  %%CITATION = PHRVA,D73,064016;%%


%\cite{Atick:1988si}
\bibitem{Atick:1988si}
  J.~J.~Atick and E.~Witten,
  ``The Hagedorn Transition And The Number Of Degrees Of Freedom Of String
  Theory,''
  Nucl.\ Phys.\ B {\bf 310} (1988) 291.
  %%CITATION = NUPHA,B310,291;%%

%\cite{Adams:2001sv}
\bibitem{Adams:2001sv}
  A.~Adams, J.~Polchinski and E.~Silverstein,
  ``Don't panic! Closed string tachyons in ALE space-times,''
  JHEP {\bf 0110}, 029 (2001)
  [arXiv:hep-th/0108075].
  %%CITATION = JHEPA,0110,029;%%

%\cite{Adams:2005rb}
\bibitem{Adams:2005rb}
  A.~Adams, X.~Liu, J.~McGreevy, A.~Saltman and E.~Silverstein,
  ``Things fall apart: Topology change from winding tachyons,''
  JHEP {\bf 0510}, 033 (2005)
  [arXiv:hep-th/0502021].
  %%CITATION = JHEPA,0510,033;%%

%\cite{Silverstein:2006tm}
\bibitem{Silverstein:2006tm}
  E.~Silverstein,
  ``Singularities and closed string tachyons,''
  arXiv:hep-th/0602230.
  %%CITATION = HEP-TH/0602230;%%

%\cite{Horowitz:2005vp}
\bibitem{Horowitz:2005vp}
  G.~T.~Horowitz,
  ``Tachyon condensation and black strings,''
  JHEP {\bf 0508}, 091 (2005)
  [arXiv:hep-th/0506166].
  %%CITATION = JHEPA,0508,091;%%


%\cite{Headrick:2006ti}
\bibitem{Headrick:2006ti}
  M.~Headrick and T.~Wiseman,
  ``Ricci flow and black holes,''
  Class.\ Quant.\ Grav.\  {\bf 23}, 6683 (2006)
  [arXiv:hep-th/0606086].
  %%CITATION = CQGRD,23,6683;%%

%\cite{Harvey:2001wm}
\bibitem{Harvey:2001wm}
  J.~A.~Harvey, D.~Kutasov, E.~J.~Martinec and G.~W.~Moore,
  ``Localized tachyons and RG flows,''
  arXiv:hep-th/0111154.
  %%CITATION = HEP-TH/0111154;%%

\bibitem{localized tachyons}
  C.~Vafa,
  %``Mirror symmetry and closed string tachyon condensation,''
  arXiv:hep-th/0111051.
  %%CITATION = HEP-TH/0111051;%%
  M.~Headrick,
  %``Decay of C/Z(n): Exact supergravity solutions,''
  JHEP {\bf 0403}, 025 (2004)
  [arXiv:hep-th/0312213].
  %%CITATION = JHEPA,0403,025;%%
  Y.~Okawa and B.~Zwiebach,
  %``Twisted tachyon condensation in closed string field theory,''
  JHEP {\bf 0403}, 056 (2004)
  [arXiv:hep-th/0403051].
  %%CITATION = JHEPA,0403,056;%%
  M.~Headrick, S.~Minwalla and T.~Takayanagi,
  %``Closed string tachyon condensation: An overview,''
  Class.\ Quant.\ Grav.\  {\bf 21}, S1539 (2004)
  [arXiv:hep-th/0405064].
  %%CITATION = CQGRD,21,S1539;%%
  O.~Bergman and S.~S.~Razamat,
  %``On the CSFT approach to localized closed string tachyons,''
  JHEP {\bf 0501}, 014 (2005)
  [arXiv:hep-th/0410046].
  %%CITATION = JHEPA,0501,014;%%
  A.~Adams, X.~Liu, J.~McGreevy, A.~Saltman and E.~Silverstein,
  %``Things fall apart: Topology change from winding tachyons,''
  JHEP {\bf 0510}, 033 (2005)
  [arXiv:hep-th/0502021].
  %%CITATION = JHEPA,0510,033;%%
  G.~T.~Horowitz,
  %``Tachyon condensation and black strings,''
  JHEP {\bf 0508}, 091 (2005)
  [arXiv:hep-th/0506166].
  %%CITATION = JHEPA,0508,091;%%
  S.~F.~Ross,
  %``Winding tachyons in asymptotically supersymmetric black strings,''
  JHEP {\bf 0510}, 112 (2005)
  [arXiv:hep-th/0509066].
  %%CITATION = JHEPA,0510,112;%%
  O.~Bergman and S.~Hirano,
  %``Semi-localized instability of the Kaluza-Klein linear dilaton vacuum,''
  Nucl.\ Phys.\  B {\bf 744}, 136 (2006)
  [arXiv:hep-th/0510076].
  %%CITATION = NUPHA,B744,136;%%

%\cite{Zamolodchikov:1986gt}
\bibitem{Zamolodchikov:1986gt}
  A.~B.~Zamolodchikov,
  ``Irreversibility of the Flux of the Renormalization Group in a 2D Field
  %Theory,''
  JETP Lett.\  {\bf 43}, 730 (1986)
  [Pisma Zh.\ Eksp.\ Teor.\ Fiz.\  {\bf 43}, 565 (1986)].
  %%CITATION = ZFPRA,43,565;%%

%\cite{Berkooz:2007nm}
\bibitem{Berkooz:2007nm}
  M.~Berkooz and D.~Reichmann,
  ``A short review of time dependent solutions and space-like singularities in
  string theory,''
  arXiv:0705.2146 [hep-th].
  %%CITATION = ARXIV:0705.2146;%%

\bibitem{applicationtimedep}
  J.~McGreevy and E.~Silverstein,
  %``The tachyon at the end of the universe,''
  JHEP {\bf 0508}, 090 (2005)
  [arXiv:hep-th/0506130].
  %%CITATION = JHEPA,0508,090;%%
  M.~Berkooz, Z.~Komargodski, D.~Reichmann and V.~Shpitalnik,
  %``Flow of geometries and instantons on the null orbifold,''
  JHEP {\bf 0512}, 018 (2005)
  [arXiv:hep-th/0507067].
  %%CITATION = JHEPA,0512,018;%%
   J.~H.~She,
  %``A matrix model for Misner universe,''
  JHEP {\bf 0601}, 002 (2006)
  [arXiv:hep-th/0509067].
  %%CITATION = JHEPA,0601,002;%%
  E.~Silverstein,
  %``Dimensional mutation and spacelike singularities,''
  Phys.\ Rev.\  D {\bf 73}, 086004 (2006)
  [arXiv:hep-th/0510044].
  %%CITATION = PHRVA,D73,086004;%%
  Y.~Hikida and T.~S.~Tai,
  %``D-instantons and closed string tachyons in Misner space,''
  JHEP {\bf 0601}, 054 (2006)
  [arXiv:hep-th/0510129].
  %%CITATION = JHEPA,0601,054;%%
   J.~H.~She,
  %``Winding string condensation and noncommutative deformation of spacelike
  %singularity,''
  Phys.\ Rev.\  D {\bf 74}, 046005 (2006)
  [arXiv:hep-th/0512299].
  %%CITATION = PHRVA,D74,046005;%%
 Y.~Nakayama, S.~J.~Rey and Y.~Sugawara,
  %``The nothing at the beginning of the universe made precise,''
  arXiv:hep-th/0606127.
  %%CITATION = HEP-TH/0606127;%%
  Y.~Hikida,
  %``Interactions for winding strings in Misner space,''
  Phys.\ Rev.\  D {\bf 75}, 046002 (2007)
  [arXiv:hep-th/0606191].
  %%CITATION = PHRVA,D75,046002;%%

%\cite{Balasubramanian:2005bg}
\bibitem{Balasubramanian:2005bg}
  V.~Balasubramanian, K.~Larjo and J.~Simon,
  ``Much ado about nothing,''
  Class.\ Quant.\ Grav.\  {\bf 22} (2005) 4149
  [arXiv:hep-th/0502111].
  %%CITATION = CQGRD,22,4149;%%

%\cite{He:2007ji}
\bibitem{He:2007ji}
  J.~He and M.~Rozali,
  ``On Bubbles of Nothing in AdS/CFT,''
  arXiv:hep-th/0703220.
  %%CITATION = HEP-TH/0703220;%%

%\cite{Zamolodchikov:1986gt}
\bibitem{Zamolodchikov:1986gt}
  A.~B.~Zamolodchikov,
  ``Irreversibility of the Flux of the Renormalization Group in a 2D Field Theory,''
  JETP Lett.\  {\bf 43}, 730 (1986)
  [Pisma Zh.\ Eksp.\ Teor.\ Fiz.\  {\bf 43}, 565 (1986)].
  %%CITATION = ZFPRA,43,565;%%

\bibitem{bulk tachyons}
  B.~Zwiebach,
  %``A solvable toy model for tachyon condensation in string field theory,''
  JHEP {\bf 0009}, 028 (2000)
  [arXiv:hep-th/0008227].
  %%CITATION = JHEPA,0009,028;%%
  D.~Kutasov, M.~Marino and G.~W.~Moore,
  %``Some exact results on tachyon condensation in string field theory,''
  JHEP {\bf 0010}, 045 (2000)
  [arXiv:hep-th/0009148].
  %%CITATION = JHEPA,0010,045;%%
  H.~Yang and B.~Zwiebach,
  %``A closed string tachyon vacuum?,''
  JHEP {\bf 0509}, 054 (2005)
  [arXiv:hep-th/0506077].
  %%CITATION = JHEPA,0509,054;%%
  O.~Bergman and S.~S.~Razamat,
  %``Toy models for closed string tachyon solitons,''
  JHEP {\bf 0611}, 063 (2006)
  [arXiv:hep-th/0607037].
  %%CITATION = JHEPA,0611,063;%%
  S.~Hellerman and I.~Swanson,
  %``Cosmological solutions of supercritical string theory,''
  arXiv:hep-th/0611317.
  %%CITATION = HEP-TH/0611317;%%
  O.~Aharony and E.~Silverstein,
  %``Supercritical stability, transitions and (pseudo)tachyons,''
  Phys.\ Rev.\  D {\bf 75}, 046003 (2007)
  [arXiv:hep-th/0612031].
  %%CITATION = PHRVA,D75,046003;%%
  S.~Hellerman and I.~Swanson,
  %``Dimension-changing exact solutions of string theory,''
  arXiv:hep-th/0612051.
  %%CITATION = HEP-TH/0612051;%%
  S.~Hellerman and I.~Swanson,
  %``Cosmological unification of string theories,''
  arXiv:hep-th/0612116.
  %%CITATION = HEP-TH/0612116;%%

%\cite{FZZ}
\bibitem{FZZ}
  V.~Fateev, A.~Zamolodchikov and Al.~Zamolodchikov, unpublished

%\cite{Lin:2007gi}
\bibitem{Lin:2007gi}
  F.~L.~Lin, T.~Matsuo and D.~Tomino,
  ``Hagedorn Strings and Correspondence Principle in AdS(3),''
  arXiv:0705.4514 [hep-th].
  %%CITATION = ARXIV:0705.4514;%%

%\cite{Rangamani:2007ju}
\bibitem{Rangamani:2007ju}
    M.~Rangamani and S.~F.~Ross,
    ''Winding tachyons in BTZ,''
    arXiv:0706.0663 [hep-th].
  %%CITATION = ARXIV:0706.0663;%%

%\cite{Banados:1992wn}
\bibitem{Banados:1992wn}
  M.~Banados, C.~Teitelboim and J.~Zanelli,
  ``The Black hole in three-dimensional space-time,''
  Phys.\ Rev.\ Lett.\  {\bf 69}, 1849 (1992)
  [arXiv:hep-th/9204099].
  %%CITATION = PRLTA,69,1849;%%

%\cite{Banados:1992gq}
\bibitem{Banados:1992gq}
  M.~Banados, M.~Henneaux, C.~Teitelboim and J.~Zanelli,
  ``Geometry of the (2+1) black hole,''
  Phys.\ Rev.\  D {\bf 48} (1993) 1506
  [arXiv:gr-qc/9302012].
  %%CITATION = PHRVA,D48,1506;%%

%\cite{Maldacena:1998bw}
\bibitem{Maldacena:1998bw}
  J.~M.~Maldacena and A.~Strominger,
  ``AdS(3) black holes and a stringy exclusion principle,''
  JHEP {\bf 9812} (1998) 005
  [arXiv:hep-th/9804085].
  %%CITATION = HEP-TH 9804085;%%

%\cite{Dijkgraaf:2000fq}
\bibitem{Dijkgraaf:2000fq}
  R.~Dijkgraaf, J.~M.~Maldacena, G.~W.~Moore and E.~P.~Verlinde,
  ``A black hole farey tail,''
  arXiv:hep-th/0005003.
  %%CITATION = HEP-TH/0005003;%%

%\cite{deBoer:2006vg}
\bibitem{deBoer:2006vg}
  J.~de Boer, M.~C.~N.~Cheng, R.~Dijkgraaf, J.~Manschot and E.~Verlinde,
  ``A farey tail for attractor black holes,''
  JHEP {\bf 0611} (2006) 024
  [arXiv:hep-th/0608059].
  %%CITATION = JHEPA,0611,024;%%

%\cite{Maldacena:2000hw}
\bibitem{Maldacena:2000hw}
  J.~M.~Maldacena and H.~Ooguri,
  ``Strings in AdS(3) and SL(2,R) WZW model. I,''
  J.\ Math.\ Phys.\  {\bf 42} (2001) 2929
  [arXiv:hep-th/0001053].
  %%CITATION = HEP-TH 0001053;%%

%\cite{Maldacena:2000kv}
\bibitem{Maldacena:2000kv}
  J.~M.~Maldacena, H.~Ooguri and J.~Son,
  ``Strings in AdS(3) and the SL(2,R) WZW model. II: Euclidean black hole,''
  J.\ Math.\ Phys.\  {\bf 42} (2001) 2961
  [arXiv:hep-th/0005183].
  %%CITATION = HEP-TH 0005183;%%

%\cite{Maldacena:2001km}
\bibitem{Maldacena:2001km}
  J.~M.~Maldacena and H.~Ooguri,
  ``Strings in AdS(3) and the SL(2,R) WZW model. III: Correlation  functions,''
  Phys.\ Rev.\ D {\bf 65} (2002) 106006
  [arXiv:hep-th/0111180].
  %%CITATION = HEP-TH 0111180;%%

%\cite{Giveon:2005mi}
\bibitem{Giveon:2005mi}
  A.~Giveon, D.~Kutasov, E.~Rabinovici and A.~Sever,
  ``Phases of quantum gravity in AdS(3) and linear dilaton backgrounds,''
  Nucl.\ Phys.\ B {\bf 719}, 3 (2005)
  [arXiv:hep-th/0503121].
  %%CITATION = HEP-TH 0503121;%%

%\cite{Gawedzki:1991yu}
\bibitem{Gawedzki:1991yu}
  K.~Gawedzki,
  ``Noncompact WZW conformal field theories,''
  arXiv:hep-th/9110076.
  %%CITATION = HEP-TH/9110076;%%


%\cite{Aharony:1999ti}
\bibitem{Aharony:1999ti}
  O.~Aharony, S.~S.~Gubser, J.~M.~Maldacena, H.~Ooguri and Y.~Oz,
  ``Large N field theories, string theory and gravity,''
  Phys.\ Rept.\  {\bf 323}, 183 (2000)
  [arXiv:hep-th/9905111].
  %%CITATION = PRPLC,323,183;%%

%\cite{Seiberg:1999xz}
\bibitem{Seiberg:1999xz}
  N.~Seiberg and E.~Witten,
  ``The D1/D5 system and singular CFT,''
  JHEP {\bf 9904} (1999) 017
  [arXiv:hep-th/9903224].
  %%CITATION = JHEPA,9904,017;%%

%\cite{Kutasov:1999xu}
\bibitem{Kutasov:1999xu}
  D.~Kutasov and N.~Seiberg,
  ``More comments on string theory on AdS(3),''
  JHEP {\bf 9904}, 008 (1999) [arXiv:hep-th/9903219].
  %%CITATION = HEP-TH 9903219;%%


%\cite{Giveon:2001up}
\bibitem{Giveon:2001up}
  A.~Giveon and D.~Kutasov,
  ``Notes on AdS(3),''
  Nucl.\ Phys.\  B {\bf 621}, 303 (2002)
  [arXiv:hep-th/0106004].
  %%CITATION = NUPHA,B621,303;%%

%\cite{Maldacena:1999mh}
\bibitem{Maldacena:1999mh}
  J.~M.~Maldacena and J.~G.~Russo,
  ``Large N limit of non-commutative gauge theories,''
  JHEP {\bf 9909} (1999) 025
  [arXiv:hep-th/9908134].
  %%CITATION = JHEPA,9909,025;%%

%\cite{Dhar:1999ax}
\bibitem{Dhar:1999ax}
  A.~Dhar, G.~Mandal, S.~R.~Wadia and K.~P.~Yogendran,
  ``D1/D5 system with B-field, noncommutative geometry and the CFT of the
  Higgs branch,''
  Nucl.\ Phys.\  B {\bf 575} (2000) 177
  [arXiv:hep-th/9910194].
  %%CITATION = NUPHA,B575,177;%%

%\cite{Polchinski:1985zf}
\bibitem{Polchinski:1985zf}
  J.~Polchinski,
  ``Evaluation Of The One Loop String Path Integral,''
  Commun.\ Math.\ Phys.\  {\bf 104} (1986) 37.
  %%CITATION = CMPHA,104,37;%%

%\cite{Berkooz:1997cq}
\bibitem{Berkooz:1997cq}
  M.~Berkooz, M.~Rozali and N.~Seiberg,
  ``Matrix description of M theory on T**4 and T**5,''
  Phys.\ Lett.\  B {\bf 408} (1997) 105
  [arXiv:hep-th/9704089].
  %%CITATION = PHLTA,B408,105;%%

%\cite{Seiberg:1997zk}
\bibitem{Seiberg:1997zk}
  N.~Seiberg,
  ``New theories in six dimensions and matrix description of M-theory on  T**5
  and T**5/Z(2),''
  Phys.\ Lett.\  B {\bf 408} (1997) 98
  [arXiv:hep-th/9705221].
  %%CITATION = PHLTA,B408,98;%%

%\cite{Giveon:1999px}
\bibitem{Giveon:1999px}
  A.~Giveon and D.~Kutasov,
  ``Little string theory in a double scaling limit,''
  JHEP {\bf 9910}, 034 (1999)
  [arXiv:hep-th/9909110].
  %%CITATION = JHEPA,9910,034;%%

%\cite{Aharony:1998ub}
\bibitem{Aharony:1998ub}
  O.~Aharony, M.~Berkooz, D.~Kutasov and N.~Seiberg,
  ``Linear dilatons, NS5-branes and holography,''
  JHEP {\bf 9810} (1998) 004
  [arXiv:hep-th/9808149].
  %%CITATION = JHEPA,9810,004;%%

%\cite{Kazakov:2000pm}
\bibitem{Kazakov:2000pm}
  V.~Kazakov, I.~K.~Kostov and D.~Kutasov,
  ``A matrix model for the two-dimensional black hole,''
  Nucl.\ Phys.\  B {\bf 622} (2002) 141
  [arXiv:hep-th/0101011].
  %%CITATION = NUPHA,B622,141;%%

%\cite{Aharony:2004xn}
\bibitem{Aharony:2004xn}
  O.~Aharony, A.~Giveon and D.~Kutasov,
  ``LSZ in LST,''
  Nucl.\ Phys.\  B {\bf 691} (2004) 3
  [arXiv:hep-th/0404016].
  %%CITATION = NUPHA,B691,3;%%

%\cite{Kim:2005av}
\bibitem{Kim:2005av}
  J.~Kim, B.~H.~Lee, C.~Park and C.~Rim,
  ``Two point correlation function of sine-Liouville theory,''
  J.\ Korean Phys.\ Soc.\  {\bf 46}, 1311 (2005)
  [arXiv:hep-th/0503050].
  %%CITATION = JKPSD,46,1311;%%

%\cite{Fukuda:2001jd}
\bibitem{Fukuda:2001jd}
  T.~Fukuda and K.~Hosomichi,
  ``Three-point functions in sine-Liouville theory,''
  JHEP {\bf 0109}, 003 (2001)
  [arXiv:hep-th/0105217].
  %%CITATION = JHEPA,0109,003;%%

%\cite{Witten:1991yr}
\bibitem{Witten:1991yr}
  E.~Witten,
  ``On string theory and black holes,''
  Phys.\ Rev.\  D {\bf 44} (1991) 314.
  %%CITATION = PHRVA,D44,314;%%

%\cite{Mandal:1991tz}
\bibitem{Mandal:1991tz}
  G.~Mandal, A.~M.~Sengupta and S.~R.~Wadia,
  ``Classical Solutions Of Two-Dimensional String Theory,''
  Mod.\ Phys.\ Lett.\  A {\bf 6} (1991) 1685.
  %%CITATION = MPLAE,A6,1685;%%

%\cite{Elitzur:1991cb}
\bibitem{Elitzur:1991cb}
  S.~Elitzur, A.~Forge and E.~Rabinovici,
  ``Some global aspects of string compactifications,''
  Nucl.\ Phys.\  B {\bf 359}, 581 (1991).
  %%CITATION = NUPHA,B359,581;%%

%\cite{Dijkgraaf:1991ba}
\bibitem{Dijkgraaf:1991ba}
  R.~Dijkgraaf, H.~L.~Verlinde and E.~P.~Verlinde,
  ``String propagation in a black hole geometry,''
  Nucl.\ Phys.\  B {\bf 371}, 269 (1992).
  %%CITATION = NUPHA,B371,269;%%

\cite{Knizhnik:1988ak}
\bibitem{Knizhnik:1988ak}
  V.~G.~Knizhnik, A.~M.~Polyakov and A.~B.~Zamolodchikov,
  ``Fractal structure of 2d-quantum gravity,''
  Mod.\ Phys.\ Lett.\  A {\bf 3}, 819 (1988).
  %%CITATION = MPLAE,A3,819;%%

%\cite{Giribet:2007uh}
\bibitem{Giribet:2007uh}
  G.~Giribet and M.~Leoni,
  ``A twisted FZZ-like dual for the 2D black hole,''
  arXiv:0706.0036 [hep-th].
  %%CITATION = ARXIV:0706.0036;%%

%\cite{Hori:2001ax}
\bibitem{Hori:2001ax}
  K.~Hori and A.~Kapustin,
  ``Duality of the fermionic 2d black hole and N = 2 Liouville theory as mirror symmetry,''
  JHEP {\bf 0108}, 045 (2001)
  [arXiv:hep-th/0104202].
  %%CITATION = JHEPA,0108,045;%%

%\cite{Giveon:2003wn}
\bibitem{Giveon:2003wn}
  A.~Giveon, A.~Konechny, A.~Pakman and A.~Sever,
  ``Type 0 strings in a 2-d black hole,''
  JHEP {\bf 0310} (2003) 025
  [arXiv:hep-th/0309056].
  %%CITATION = JHEPA,0310,025;%%

%\cite{Israel:2004jt}
\bibitem{Israel:2004jt}
  D.~Israel, A.~Pakman and J.~Troost,
  %``D-branes in N = 2 Liouville theory and its mirror,''
  Nucl.\ Phys.\  B {\bf 710} (2005) 529
  [arXiv:hep-th/0405259].
  %%CITATION = NUPHA,B710,529;%%

%\cite{Maldacena:2005hi}
\bibitem{Maldacena:2005hi}
  J.~M.~Maldacena,
  ``Long strings in two dimensional string theory and non-singlets in the
  matrix model,''
  JHEP {\bf 0509} (2005) 078
  [Int.\ J.\ Geom.\ Meth.\ Mod.\ Phys.\  {\bf 3} (2006) 1]
  [arXiv:hep-th/0503112].
  %%CITATION = 00436,3,1;%%

%\cite{Giveon:1994fu}
\bibitem{Giveon:1994fu}
  A.~Giveon, M.~Porrati and E.~Rabinovici,
  ``Target space duality in string theory,''
  Phys.\ Rept.\  {\bf 244} (1994) 77
  [arXiv:hep-th/9401139].
  %%CITATION = HEP-TH 9401139;%%

%\cite{Horne:1991gn}
\bibitem{Horne:1991gn}
  J.~H.~Horne and G.~T.~Horowitz,
  ``Exact black string solutions in three-dimensions,''
  Nucl.\ Phys.\ B {\bf 368} (1992) 444
  [arXiv:hep-th/9108001].
  %%CITATION = HEP-TH 9108001;%%

%\cite{Giveon:1998ns}
\bibitem{Giveon:1998ns}
  A.~Giveon, D.~Kutasov and N.~Seiberg,
  ``Comments on string theory on AdS(3),''
  Adv.\ Theor.\ Math.\ Phys.\  {\bf 2}, 733 (1998)
  [arXiv:hep-th/9806194].
  %%CITATION = HEP-TH 9806194;%%

%\cite{Fukuma:1999jt}
\bibitem{Fukuma:1999jt}
  M.~Fukuma, T.~Oota and H.~Tanaka,
  ``Comments on T-dualities of Ramond-Ramond potentials on tori,''
  Prog.\ Theor.\ Phys.\  {\bf 103} (2000) 425
  [arXiv:hep-th/9907132].
  %%CITATION = PTPKA,103,425;%%


%\cite{Hassan:1999bv}
\bibitem{Hassan:1999bv}
  S.~F.~Hassan,
  ``T-duality, space-time spinors and R-R fields in curved backgrounds,''
  Nucl.\ Phys.\  B {\bf 568} (2000) 145
  [arXiv:hep-th/9907152].
  %%CITATION = NUPHA,B568,145;%%

%\cite{Giveon:2006pr}
\bibitem{Giveon:2006pr}
  A.~Giveon and D.~Kutasov,
  ``Fundamental strings and black holes,''
  JHEP {\bf 0701}, 071 (2007)
  [arXiv:hep-th/0611062].
  %%CITATION = JHEPA,0701,071;%%

%\cite{Horowitz:1997jc}
\bibitem{Horowitz:1997jc}
  G.~T.~Horowitz and J.~Polchinski,
  ``Self gravitating fundamental strings,''
  Phys.\ Rev.\  D {\bf 57}, 2557 (1998)
  [arXiv:hep-th/9707170].
  %%CITATION = PHRVA,D57,2557;%%

%\cite{Horowitz:1996nw}
\bibitem{Horowitz:1996nw}
  G.~T.~Horowitz and J.~Polchinski,
  ``A correspondence principle for black holes and strings,''
  Phys.\ Rev.\  D {\bf 55}, 6189 (1997)
  [arXiv:hep-th/9612146].
  %%CITATION = PHRVA,D55,6189;%%

%\cite{Kurita:2004yn}
\bibitem{Kurita:2004yn}
  Y.~Kurita and M.~a.~Sakagami,
  ``CFT description of three-dimensional Hawking-Page transition,''
  Prog.\ Theor.\ Phys.\  {\bf 113}, 1193 (2005)
  [arXiv:hep-th/0403091].
  %%CITATION = PTPKA,113,1193;%%

%\cite{Polchinski:1998rq}
\bibitem{Polchinski:1998rq}
  J.~Polchinski,
  ``String theory. Vol. 1: An introduction to the bosonic string,''
  \href{http://www.slac.stanford.edu/spires/find/hep/www?irn=4634799}{SPIRES entry}
  {\it  Cambridge, UK: Univ. Pr. (1998) 402 p}

%\cite{Cornalba:2002fi}
\bibitem{Cornalba:2002fi}
  L.~Cornalba and M.~S.~Costa,
  ``A new cosmological scenario in string theory,''
  Phys.\ Rev.\  D {\bf 66} (2002) 066001
  [arXiv:hep-th/0203031].
  %%CITATION = PHRVA,D66,066001;%%
  L.~Cornalba and M.~S.~Costa,
  ``Time-dependent orbifolds and string cosmology,''
  Fortsch.\ Phys.\  {\bf 52} (2004) 145
  [arXiv:hep-th/0310099].
  %%CITATION = FPYKA,52,145;%%


%\cite{Pioline:2003bs}
\bibitem{Pioline:2003bs}
  B.~Pioline and M.~Berkooz,
  ``Strings in an electric field, and the Milne universe,''
  JCAP {\bf 0311}, 007 (2003)
  [arXiv:hep-th/0307280].
  %%CITATION = JCAPA,0311,007;%%

%\cite{Berkooz:2002je}
\bibitem{Berkooz:2002je}
  M.~Berkooz, B.~Craps, D.~Kutasov and G.~Rajesh,
  ``Comments on cosmological singularities in string theory,''
  JHEP {\bf 0303}, 031 (2003)
  [arXiv:hep-th/0212215].
  %%CITATION = JHEPA,0303,031;%%


\end{thebibliography}
\end{document}